\documentclass[pra,twocolumn,nofootinbib,superscriptaddress,amssymb]{revtex4}
\usepackage[usenames,dvipsnames]{color}
\usepackage{psfrag,graphicx}
\usepackage{dcolumn}
\usepackage{bm}
\usepackage{amsfonts,amssymb,amsmath}
\usepackage{tabularx, booktabs}
\usepackage{slashed}
\usepackage{makecell}
\usepackage{scrextend}
\usepackage{diagbox}
\usepackage[colorlinks]{hyperref}

\newcommand\blfootnote[1]{%
  \begingroup
  \renewcommand\thefootnote{}\footnote{#1}%
  \addtocounter{footnote}{-1}%
  \endgroup
}
\hypersetup{
plainpages=true,
breaklinks=true,
hypertexnames=false,
 pageanchor=true,
colorlinks=true,
linkcolor={blue},
citecolor={magenta},
urlcolor={blue},
anchorcolor={black}
}
\newcolumntype{b}{X}
\newcolumntype{s}{>{\hsize=.08\hsize}X}
\newcolumntype{t}{>{\hsize=.05\hsize}X}
\newcolumntype{L}{>{\raggedright\arraybackslash}X} % for ragged-right material
\newcolumntype{C}{>{\centering\arraybackslash}X}  % for centered material
\newcommand{\be}{\begin{equation}}
\newcommand{\ee}{\end{equation}}
\newcommand{\bq}{\begin{eqnarray}}
\newcommand{\eq}{\end{eqnarray}}

\newcommand{\ket}[1]{ |#1 \rangle}
\newcommand{\bra}[1]{ \langle #1  |}

\newcommand{\Eq}[1]{Eq.~(\ref{#1})}

\newenvironment{red}{\color{red}}

\newenvironment{blue}{\color{blue}}

\usepackage{circuitikz}
\usepackage{color}
\usepackage{graphicx}
\usepackage{epstopdf}

\usepackage{tikz}
\usetikzlibrary[shapes.arrows]
\usetikzlibrary{shapes}
\usetikzlibrary{arrows, decorations.markings}
\tikzstyle{vecArrow} = [thick, decoration={markings,mark=at position
  1 with {\arrow[semithick]{open triangle 60}}},
  double distance=1.4pt, shorten >= 5.5pt,
  preaction = {decorate},
  postaction = {draw,line width=1.4pt, white,shorten >= 4.5pt}]
\tikzstyle{innerWhite} = [semithick, white,line width=1.4pt, shorten >= 4.5pt]

\usepgflibrary{arrows}
\usetikzlibrary{arrows}
\usetikzlibrary{positioning}
\usetikzlibrary{calc}
\usetikzlibrary{decorations.pathmorphing}
\usetikzlibrary{backgrounds}

\definecolor{black}{RGB}{0,0,255}
\definecolor{RedOrange}{RGB}{255,165,0}
\definecolor{DarkGreen}{RGB}{0,150,0}
\definecolor{Green}{RGB}{0,210,0}

\definecolor{RedOrange1}{RGB}{255,140,0}
\definecolor{RedOrange2}{RGB}{255,69,0}
\definecolor{black1}{RGB}{0,0,205}
\definecolor{black2}{RGB}{0,191,255}
\definecolor{fuchsia}{RGB}{152,59,192}

\definecolor{YS}{rgb}{153,153,0}

\begin{document}
\title{Multielectron Ground State Electroluminescence}

\author{Mauro Cirio}
\thanks{These two authors contributed equally to this work.\\
cirio.mauro@gmail.com\\
nathan.shammah@gmail.com
}
\affiliation{Graduate School of China Academy of Engineering Physics, Beijing 100193, China}
\affiliation{Theoretical Quantum Physics Laboratory, RIKEN Cluster for Pioneering Research, Wako-shi, Saitama 351-0198, Japan}

\author{Nathan Shammah}
\thanks{These two authors contributed equally to this work.\\
cirio.mauro@gmail.com\\
nathan.shammah@gmail.com
}
\affiliation{Theoretical Quantum Physics Laboratory, RIKEN Cluster for Pioneering Research, Wako-shi, Saitama 351-0198, Japan}

\author{Neill Lambert}
\affiliation{Theoretical Quantum Physics Laboratory, RIKEN Cluster for Pioneering Research, Wako-shi, Saitama 351-0198, Japan}

\author{Simone De Liberato}
\affiliation{School of Physics and Astronomy, University of Southampton, Southampton SO17 1BJ, United Kingdom}

\author{Franco Nori}
\affiliation{Theoretical Quantum Physics Laboratory, RIKEN Cluster for Pioneering Research, Wako-shi, Saitama 351-0198, Japan}
\affiliation{Department of Physics, University of Michigan, Ann Arbor, MI 48109-1040, USA}

\date{\today}
%\pacs{}
\begin{abstract}
The ground state of a cavity-electron system in the ultrastrong coupling regime is characterized by the presence of virtual photons. 
If an electric current flows through this system, the modulation of the light-matter coupling induced by this non-equilibrium effect can induce an extra-cavity photon emission signal, even when electrons entering the cavity do not have enough energy to populate the excited states. 
We show that this ground-state electroluminescence, previously identified in a single-qubit system [Phys. Rev. Lett. 116, 113601 (2016)] can arise in a many-electron system. The collective enhancement of the light-matter coupling makes this effect, described beyond the rotating wave approximation, robust in the thermodynamic limit, allowing its observation in a broad range of physical systems, from a semiconductor heterostructure with flat-band dispersion to various implementations of the Dicke model.
\end{abstract}
\maketitle
\paragraph*{Introduction.---}
When the interaction between light and matter is stronger than the coupling to the environment, a variety of hybridization effects can be observed. In the context of cavity quantum electrodynamics, realizing this ``strong-coupling'' regime has been achieved in different ways; for example, by reducing the losses of the system \cite{Raimond01}, by enhancing the vacuum electromagnetic field in one-dimensional cavities \cite{Haroche}, by increasing the dipole moment of the atom \cite{Niemczyk10}, or by taking advantage of collective properties \cite{Geiser12b}. Building upon these strategies, it has been possible to engineer light-matter couplings up to a significant fraction of the bare energies of the bare light and matter modes themselves \cite{Ballarini19,Raimond01,Ciuti05,Anappara09,Niemczyk10,Schwartz11,Muravev11,Scalari12,Geiser12,Porer12,Scalari13,Gubbin14,Askenazi14,Gambino14,Maissen14,Goryachev14,Baust16,Lambert04,Nataf10,Baksic13,Feist15,Orgiu15,DeLiberato14,Ripoll15,Bamba15,Hutchison12,Hutchison13,Galego15,Cwik16,Cortese17}.
\begin{figure}[htb!]
\includegraphics[width=9cm]{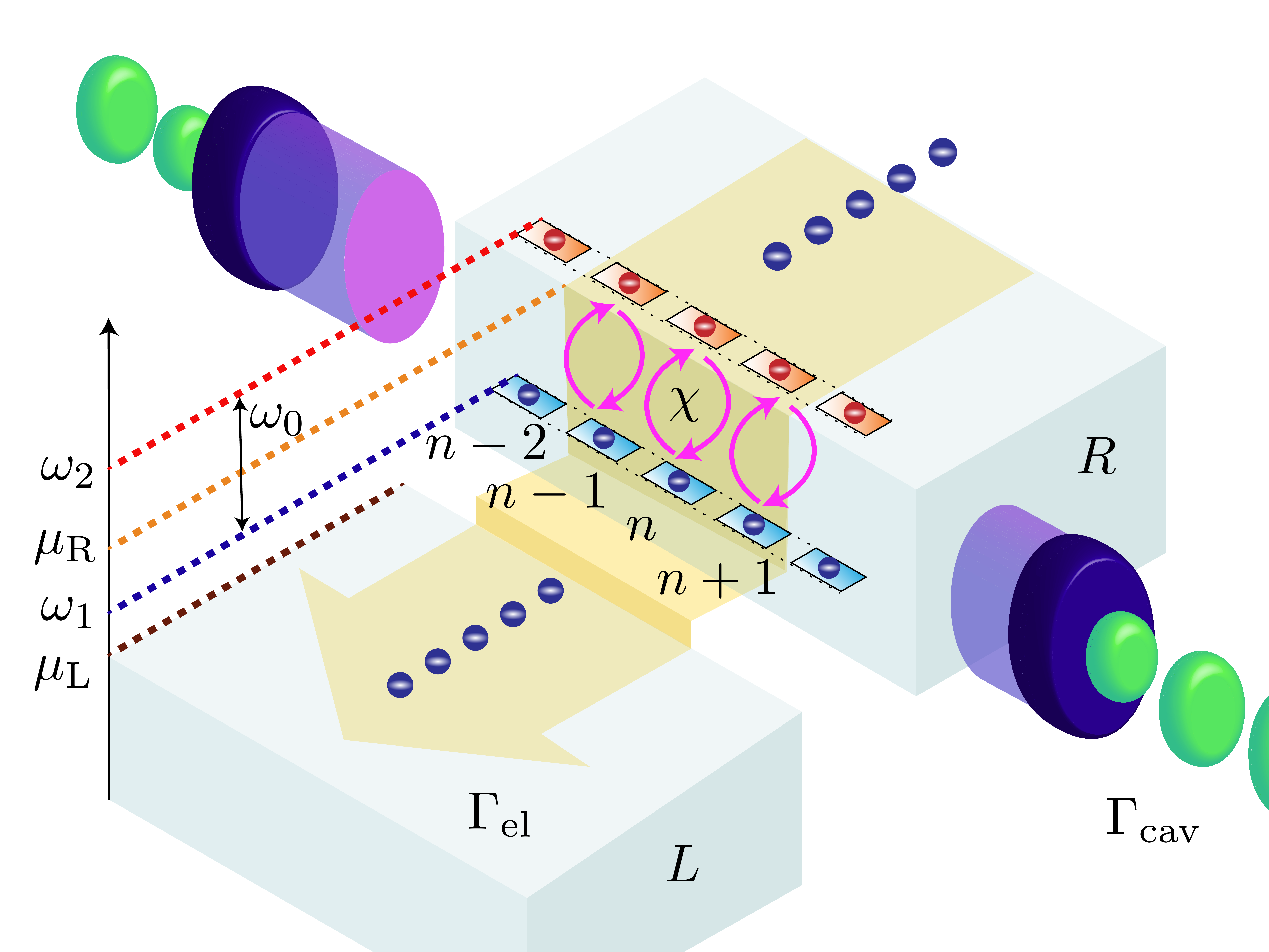}
\centering
\caption{
\label{fig}
A right lead (R) is connected to a left lead (L) via a middle region, the two elements kept at chemical potentials $\mu_R$ and $\mu_L$, respectively, by applying an electrical bias, which induces an electron current quantified by a rate $\Gamma_\text{el}$ for the free electrons (blue spheres) flowing out of the device. Sandwiched between the leads, a solid-state cavity (dark purple disks), enhances the electronic coupling to the photonic vacuum field (light purple disk cross-section), at a strength quantified by $\chi$ for each electron. 
The bare frequency difference between the two electronic flat bands, $\omega_0=\omega_2-\omega_1$, separates the lower states (blue) and upper states (red). 
The presence of virtual photons inside the cavity induces an extra-cavity photon emission (green blobs) from the polaritonic ground state, at a rate $\Gamma_\text{cav}$.}
\end{figure}

This new cavity quantum electrodynamics {(QED) ``ultra-strong'' regime has made possible the observation and study of a range of unique} physical effects \cite{Dodonov06,DeLiberato07,DeLiberato09,Agnesi09,Faccio11,Carusotto12,Auer12,Stassi13,Garziano13,Garziano14,Garziano15,Stassi16,Stassi18,Kockum18b}.
Among these phenomena are the ones originating from the hybridization of the ground state. This hybridization leads to a ground state photonic population that is sometimes called ``virtual'', as it is energetically forbidden from leaking into the environment. However, there are several proposals describing how these hybridized ground-states can be observed, typically by modulating some system parameter \cite{Stassi13,Garziano13,Garziano14,Garziano15,Stassi16,Cirio18,Stassi18}, akin to the way the dynamical Casimir effect relies on amplifying vacuum fluctuations \cite{Nation12,Lambrecht96,Johansson09,Johansson10,Johansson13b,Wilson11}.

In particular, in Ref.~\cite{Cirio16} it has been shown that the passage of an electronic current through a device where, within the device, electrons ultra-strongly couple to light in a cavity, can result in
extra-cavity emission, i.e., the conversion of virtual to real photons. In Ref.~\cite{Cirio16} such ``ground state electroluminescence'' was predicted for systems in which a \emph{single} electron at a time interacts ultra-strongly with the cavity mode  \cite{Delbecq11,Delbecq13,Viennot14,Samkharadze16,Stockklauser17,Mi17,Landig18,Mi18,Samkharadze18}.
In this Letter, we analyze ground state electroluminescence in a much more general scenario in which \emph{many} electrons at the same time are allowed to interact with the cavity mode \cite{Ritsch13,Kim13,Liu15b,Cong16,Paravicini17,Bayer18,Cuevas18,Paravicini18}. This allows for stronger effective couplings through collective effects in a solid-state device \cite{Ballarini19,DeLiberato13a,Cong16,Shammah18,Kirton18r}, as sketched in Fig.~\ref{fig}.

As we will show, one could expect the electroluminescence effect to be washed out in a system containing \emph{many} electrons, because, while the coupling is enhanced by collective effects, the conversion of virtual to real photons relies on a process where an electron leaving the system effectively changes the light-matter coupling. In this many-electron system, such an effective modulation of the light-matter coupling is  suppressed with the number of electrons, so one might expect that this negates the enhanced collective coupling.

However, surprisingly, we find here that the combination of collective coupling and the many-electron nature of the current combine to make the ground state electroluminescence macroscopically robust even in the thermodynamic limit.
The transport-induced luminescent effect can be estimated by an intuitive bosonic theory that goes beyond the rotating wave approximation (RWA) by including counter-rotating terms perturbatively. In the Supplementary Material (SM) we test this model against a full bosonic model that includes non-RWA terms non-perturbatively, and a second-quantization fermionic theory, finding excellent agreement.

\paragraph*{Light-matter system.---}
  We consider a prototypical many-body fermionic system interacting with light in a solid-state quantum device. The model system can be generalized further due to the approximations that we will make, but, for definiteness, we begin by considering two electronic bands containing a maximum of $2 N_T$ electrons, which interact with a single electromagnetic mode confined in a cavity. We further neglect electron-electron interactions, band dispersion, and higher excitations. We thus consider a two flat-band electronic model such that it can be described by the Hamiltonian ($\hbar=1$ hereafter),
\begin{eqnarray}
\label{eq:full}
H&=&\omega_{c}a^\dagger a+\displaystyle\sum_n\left(\omega_1 c_{1,n}^\dagger c_{1,n} + \omega_2 c_{2,n}^\dagger c_{2,n}\right)+D(a+a^\dagger)^2
\nonumber\\
&&+\chi(a+a^\dagger) \displaystyle\sum_{n}(c_{2,n}^\dagger c_{1,n}+c_{1,n}^\dagger c_{2,n}),
\end{eqnarray}
where $c_{1,n}$ ($c_{2,n}$) represents the annihilation operator for the $n$th ($n=1,\dots,N_T$) fermion in the first (second) state with energy $\omega_1$ ($\omega_2$).
Note that in \Eq{eq:full} we are counting each fermion over the index $n$; in several solid-state systems, this can be shown to be equivalent to a model for flat bands, as in intersubband transitions with finite real in-plane momentum \cite{Anappara09,Scalari12,Scalari13}, in the limit of small photon momentum or strong magnetic confinement. In more general contexts it may be required to include the photonic momentum, which can induce diagonal transitions \cite{DeLiberato08,DeLiberato09,DeLiberato09b,DeLiberato13b,Shammah14,Shammah15}.
The annihilation operator $a$ is associated with a cavity mode of frequency $\omega_{c}$. The light-matter interaction has strength $\chi$ and the potential energy of the electromagnetic field is proportional to the frequency $D=N\chi^2/\omega_0$, { relative to the diamagnetic term \cite{Kockum18b,DeLiberato14,Ripoll15}}, where $\omega_0=\omega_2-\omega_1$.

To begin our analysis, we divide the Hilbert space in sectors closed under the Hamiltonian evolution. They are characterized by the set of sites occupied by a single electron {$\{N\}$}, the set of sites occupied by two electrons $\{N_2\}$ and the number of photons in the cavity.

Within each of these sectors, the coherent dynamics can be described by
\begin{eqnarray}
\label{eq:simplified}
H&=&\omega_{c}a^\dagger a+\omega_0 S^3 +\chi(a+a^\dagger)(S^{-}+S^+),
\end{eqnarray}
which takes the standard form of the Dicke Hamiltonian. The interaction of a cavity mode of frequency $\omega_c$ with a matter excitation of frequency $\omega_0$, where $\omega_0=\omega_2-\omega_1$, is described beyond the RWA. Here we defined $\sigma^{-}_n=c^\dagger_{1,n}c_{2,n}$, $S^-_{n}=\sigma^-_{n}$, and $S_n^3=\sigma^z_n/{2}$, where $\sigma^\alpha_n$ is the $\alpha$-direction Pauli matrix operator. 
In \Eq{eq:simplified} we performed a fermion-to-spin transformation that, with respect to \Eq{eq:full}, involves no approximations. The parameters have been renormalized following the Bogoliubov transformation needed to reabsorb the diamagnetic term proportional to $D$ in \Eq{eq:full}  (see Sec.~II of the SM).
For the sake of generality in \Eq{eq:full} we neglect the Coulomb interaction, which would depend on microscopic details. In the case of parallel subbands a theoretical description of electron-electron interactions in the bosonic approximation, directly applicable to our approach \cite{DeLiberato12,Todorov15}, can be completely captured by a renormalization of the system transition frequency, the so-called depolarization shift \cite{Nikonov97}, and by a more complex functional dependency between the electronic operators $c$ and the collective excitation operators $S^\alpha$. While such more complex relations remain quadratic, and could thus be incorporated in our treatment, its deviations from \Eq{eq:full} scale with the ratio between the plasma frequency, $\omega_\text{p}$, and the bare excitation frequency, $\omega_0$. Equation (\ref{eq:full}) thus remains quantitatively accurate while $\omega_0\gg\omega_\text{p}$.
\paragraph*{Environment.---}
\label{sec:env}
We are interested in studying the effects of three environments on this model: a left {($L$)} and right {($R$)} electronic reservoir, which give rise to the electronic current, and the extra-cavity electromagnetic modes, into which the photons are emitted. { The total environment-system interaction Hamiltonian} is
$H^I=H^I_\text{el}+H^I_\text{cav}=H^I_{L}+H^I_{R}+H^I_\text{cav}$. Our aim is to compute the transition rates among eigenstates of the system induced by the interaction Hamiltonian, $H^I$, representing the physical interaction with the environmental degrees of freedom.
We can model the interaction with the electronic reservoirs as $H^I_{L}=\lambda\sum_{n,\zeta}[(c_{1,n}+c_{2,n})c^\dagger_{L;n,\zeta}+\text{h.c.}]$,
and identically for $H^I_{R}$ (change $L\rightarrow R$), thus assuming that the energy scale, $\lambda$, is equal for the two fermionic reservoirs.
The operators $c_{L(R);n,\zeta}$ label the annihilation operators for a fermion associated with a degree of freedom $n$ and $\zeta$ in the left (right) reservoir.

Since we are interested in strong-coupling effects between light and matter only within the system, we treat all three environments perturbatively. They only induce transitions between eigenstates of the system, as given by the Fermi golden rule. The total \emph{electron-transport rates} can be calculated by summing over single electron scattering processes, $\Gamma_\text{el}^{\alpha\rightarrow \beta}=\sum_n\Gamma_{\text{el},n}^{\alpha\rightarrow \beta}$, (see Sec.~I of the SM for details) where 
\begin{equation}
\label{eq:M}
\Gamma_{\text{el},n}^{\alpha\rightarrow \beta}\propto { \Gamma_\text{el}} | M^{n}_{\alpha\beta}|^2= { \Gamma_\text{el}}|\bra{\beta}(c_{1,n}+c_{2,n})\ket{\alpha}|^2,
\end{equation}
where $\alpha$ and $\beta$ are the initial and final states for the system, { $\Gamma_\text{el}$ is the electron tunneling rate} and $M_{\alpha\beta}^n$ provides the \emph{electron-current transition matrix element} for the $n$th electron site.

To calculate the ground-state electroluminescence rate we consider that, when $N$ electrons are in the device, and the device is in the hybridized light-matter ground state, $\ket{\alpha}=\ket{G_N}$,  an electron within the device can leave, reducing the electron number to $(N-1)$. When an electron leaves, it can, due to the ground-state light-matter hybridization, result in a transition to an excited state of the hybridized system with $(N-1)$ electrons, $\ket{\beta}=\ket{E_{N-1}}$, which contains a non-zero photonic population. We assume that the cavity loss rate $\Gamma_\text{cav}$ is much faster than the electronic rates $\Gamma_\text{el}$, such that this excited state immediately decays and emits an extra-cavity photon, decaying to the $(N-1)$ ground state, $\ket{G_{N-1}}$; this emission, arising only because the ground state itself contains photons, is the electroluminescence we want to produce. In addition, by imposing a chemical potential across the system which forbids electrons from entering directly into excited states of the coupled-system, $\mu_L<\mu_R<\omega_2$ one can suppress ``regular'' electroluminescence and ensure the observed photon emission only arises from the ground state.

Under the above assumptions ($\Gamma_\text{cav}\gg \Gamma_\text{el}$ and energetically forbidden regular electroluminescence)  the overall rate of ground-state-sourced photonic emission depends upon the \emph{electron-current transition matrix elements}, $M_{\alpha\beta}^n$ of \Eq{eq:M}. This reduces to the problem of calculating the properties of the \emph{ground state}, $\ket{G_N}$, and the various possible excited states, $\ket{E_{N-1}}$, that contribute to these transitions, and the overlap with the operators which destroy electrons. In the SM we present a fully fermionic calculation of such rates, but it is much more instructive to first consider a simpler bosonic approximation, which captures the essential physics.
\paragraph*{Bosonic approximation.---} To proceed further, we assume that thermalization effects are such that we can neglect double-occupied electron sites, $N_2\simeq0$,
and consider the following approximate Holstein-Primakoff transformation
$S_+=\sqrt{N}b^\dagger+O(|{b^\dagger b}/{\sqrt{N}}|)$, and $S_z= b^\dagger b-j_N$
in terms of an effective bosonic mode $b$. In a dilute regime in which the number of electronic excitations is much smaller than the total number of electrons, we can neglect terms of order $|b^\dagger b/\sqrt{N}|$ and rewrite \Eq{eq:simplified} as
\begin{eqnarray}
\label{eq:Hbos}
H&\simeq&\omega_c a^\dagger a+\omega_0 b^\dagger b+g_N(a b^\dagger + a^\dagger b)
+g_N(ab+a^\dagger b^\dagger)
\nonumber\\&=&
H_\text{JC}+g_N(ab+a^\dagger b^\dagger)=H_\text{JC}+V,
\end{eqnarray}
up to $\mathbb{C}$-numbers and terms of order $1/\sqrt{N}$,
and where $g_N=\sqrt N\chi$ is the \emph{bosonic} light-matter coupling.
While the \emph{full} bosonic Hamiltonian of \Eq{eq:Hbos} can be diagonalized analytically (see Secs.~II, III of the SM), to most clearly highlight the main idea behind the processes studied here, we will consider the counter-rotating term $V$ as a perturbation of the Jaynes-Cummings (JC) term, $H_\text{JC}$, and rewrite \Eq{eq:Hbos} as
$H \simeq \omega^- p_-^\dagger p_- +\omega^+ p_+^\dagger p_+ +V$,
where $p^\dagger_\pm=\alpha^\pm_a a^\dagger + \alpha^\pm_b b^\dagger$ are the polaritonic excitations of the JC part of the original Hamiltonian and where the explicit expression for the polariton energies $\omega^\pm$ and the dimensionless coefficients $\alpha^\pm_a$ and $\alpha^\pm_b$ {are} given in Sec.~III of the SM.
First-order perturbation theory in $V$ gives the following expression for the ground state and single-polariton states for the non-RWA system,
\begin{eqnarray}
\ket{G_N}\!\!&=&\!\!\ket{G_N^{(0)}}\!-\beta_{++}\ket{\!\!+\!+_N^{(0)}}\!-\beta_{+-}\ket{\!\!+\!-_N^{(0)}}\!-\beta_{--}\ket{\!\!-\!-_N^{(0)}},
\nonumber\\
\ket{\pm_N}\!\!&=&\!\!\ket{\pm_N^{(0)}}+\cdots,
\label{eqstates}
\end{eqnarray}
where we introduced the perturbative coefficients $\beta_{\pm\pm}=-\sqrt{2}g_N(\alpha^\mp_a\alpha^\mp_b)/(2\omega^\pm)$ and $\beta_{+-}=g_N(\alpha_a^+\alpha_b^-+\alpha_a^-\alpha_b^+)/(\omega^+ + \omega^-)$, that are explicitly derived in Sec.~III of the SM. For the sake of clarity, we omitted higher-order terms in the expansion, which will not contribute to our results [indicated by the suspended dots in \Eq{eqstates}].
Note that the ground state is a superposition of the Fermi sea for the system with $N$ electrons in the first band and no cavity photons, $\ket{G^{(0)}_N} = \bigotimes_{n\in \mathbf{N}}c^\dagger_{1,n}\ket{0_\text{el}}\ket{0_\text{ph}}$, where $\ket{0_\text{el}}$ and $\ket{0_\text{ph}}$ represent the electronic and photonic vacuum states, respectively, and $\mathbf{N}$ is the set of occupied sites of cardinality $N$. The double-polariton states of the unperturbed basis, similarly to all of the other excited states, can be defined by multiple applications of the JC polaritons, e.g., $\ket{\pm^{(0)}_N}=p^\dagger_\pm\ket{G^{(0)}_N}$ for the single-polariton states and $\ket{\!\pm\pm^{(0)}_N}=(p^\dagger_\pm)^2\ket{G^{(0)}_N}/\sqrt{2}$ and $\ket{\!+-^{(0)}_N}=p^\dagger_+ p^\dagger_-\ket{G^{(0)}_N}$ for the double-polariton states.
\paragraph*{Ground state electroluminescence.---} We assume the system is initially in its ground-state and, by emission of an electron, can transition to an excited state, which then decays by emitting photons, the process which constitutes ground state electroluminescence. 

Setting $\mu_L<\mu_R<\omega_2$, we obtain that $\Gamma^{G\rightarrow B}_\text{el}=\Gamma_\text{el}\delta_{G,B}$,
where $G$ labels the ground-state and $B$ labels any state (see Sec.~I of the SM for details).
 This condition ensures that the regular direct electroluminescence is energetically forbidden, and allows for the undiluted ground-state process to occur. The ground-state polariton creation leading to photon emission can be estimated as
 $ \Gamma_\text{GSE}=\sum_{E=\{\pm,\pm\pm,\pm\mp,\dots\}}\Gamma^{G\rightarrow E}_\text{el}$.

We begin by calculating the transition from the ground state, $\ket{G_N}$, to the single-polariton states, $\ket{\pm_{N-1}}$. From \Eq{eq:M}, we have $M_{G\pm}^{n}=\bra{\pm_{N-1}}(c_{1,n}+c_{2,n})\ket{G_N}$,
where the state $\ket{\pm_{N-1}}$ is the state with $(N-1)$ electrons due to the tunneling of the $n$th electron.
Here we use the perturbative expressions to expand the non-RWA contribution in these states, given in \Eq{eqstates}.
To proceed further one needs to calculate expectation values of fermionic operators onto light-matter many-body states intrinsically expressed in terms of polariton operators. This task can be crucially simplified by using Eqs.~(\ref{eq:full}),~(\ref{eq:simplified}) and the Holstein-Primakoff mapping to rewrite
$\sqrt{N}b=S_{-}
=\sum_{n}c^\dagger_{1,n}c_{2,n},$ which, using the definition of the polariton modes, immediately gives
$ [p_\pm^\dagger,c_{1,n}+c_{2,n}]=\alpha^\pm_b[b^\dagger,c_{1,n}+c_{2,n}]=-\frac{1}{\sqrt{N}}\alpha^\pm_b c_{1,n}$,
which holds up to order $1/\sqrt{N}$ and holds linearly in any of the perturbative parameters (see Sec.~III of the SM for details). We then obtain an explicit expression for the matrix elements contained in \Eq{eq:M},
$M_{G\pm}^n=(\sqrt{2}\beta_{\pm\pm}\alpha_b^\pm+\beta_{+-}\alpha_b^\mp)/\sqrt{N}$,
 
which, together with the initial working condition $\Gamma_\text{cav}\gg \Gamma_\text{el}$, allows us to estimate the photon emission rate from the ground state to the single-polariton states, $\Gamma_\text{em}^\pm$, as
$\Gamma_\text{em}^{\pm}\simeq\Gamma^{G\rightarrow\pm}_{\text{el}}
\propto\sum_n |M_{G\pm}^n|^2=
|\sqrt{2}\beta_{\pm\pm}\alpha_b^\pm+\beta_{+-}\alpha_b^\mp|^2$,
that is $\Gamma_\text{em}^{\pm}=O(\eta^2)=O(N\chi^2{ /\omega_0^2})$, where $\eta=g_N/\omega_0$.

The contributions to $ \Gamma_\text{GSE}$ from the higher-excited states (which are double-polariton states,
$\ket{G_N}\rightarrow\ket{\pm\pm_{N-1}} $ and $\ket{G_N}\rightarrow\ket{+-_{N-1}}$), are
of $O(\eta^2/N)$, as detailed in Sec.~III of the SM. Thus the dominant contribution to the ground state electroluminescence (GSE) involves the single-polariton transitions, giving the total GSE rate
$\Gamma_\text{GSE}\simeq\Gamma^{+}_\text{em}+\Gamma^{-}_\text{em}=O(\eta^2)$.
 
Remarkably, this emission is of the same order of magnitude of the one predicted in systems containing a single electron \cite{Cirio16} (but following the enhanced collective coupling rate, $\eta^2=N\chi^2{ /\omega_0^2}$).
In the single electron case  \cite{Cirio16}, the light-matter coupling was strongly modulated as the single-electron coupling was assumed to be ultra-strong, and the effective modulation of the coupling due to the emission of the electron was large.
Here instead, the tunneling of a single electron (among $N$ total) only minimally modulates the light-matter coupling, yet a \emph{collective} enhancement occurs, to ensure the same $\eta^2$ scaling. This can be interpreted as a superradiant enhancement with respect to the single-particle light-matter coupling of the fermionic system, $\chi$, and the overall large electron current.
 
% Figure 2
\begin{figure}[t]
\includegraphics[width=9cm]{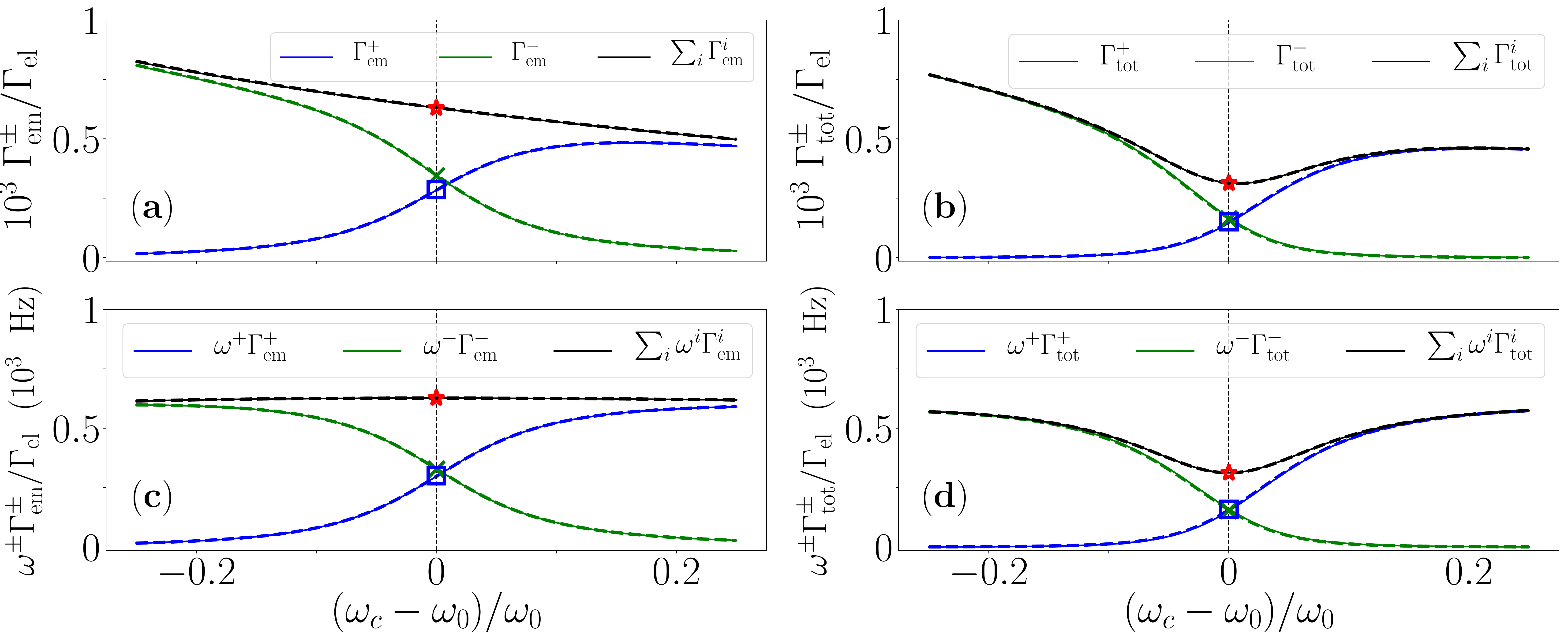}
\centering
\caption{(a,c) Polariton emission rates $\Gamma^{B'}_{\text{em}}$ and fluxes, $\omega^{B'}\Gamma^{B'}_{\text{em}}$, in units of the total electron transport rate $\Gamma_\text{el}$ for the upper polariton ($B'={+}$, blue curves) and lower polariton ($B'=-$, green curves), and sum of the two signals (black curves), versus the normalized detuning for $g _N/\omega_0=0.05$. (b,d) Total photon emission rates, from \Eq{totgammapm}, and fluxes. Solid curves correspond to the bosonic RWA quantities, dashed curves to the full boson model developed in the SM.  
\label{fig2}}
\end{figure}
% Figure 3
\begin{figure}[t]
\includegraphics[width=9cm]{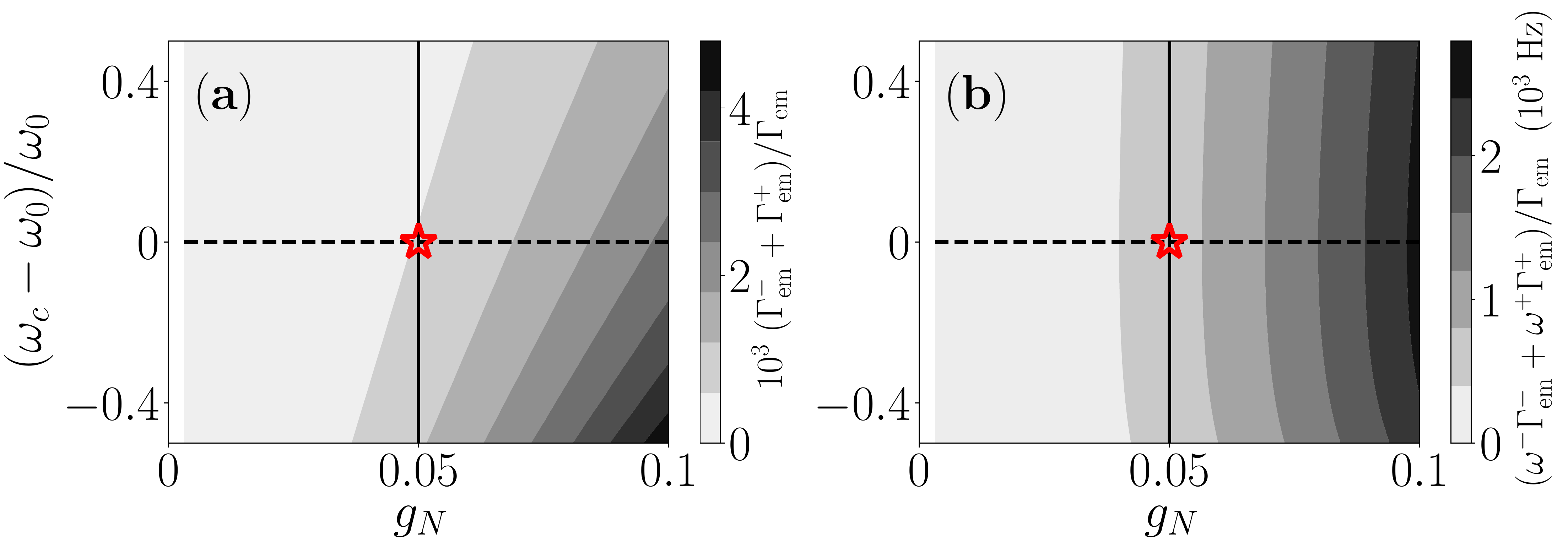}
%Code/cptotdetN
\centering
\caption{Polariton emission rate $\Gamma_{\text{em}}=\Gamma_{\text{em}}^-+\Gamma_{\text{em}}^+$ [panel (a)] and flux [panel (b)] as a function of the 
frequency detuning and coupling strength, setting $\chi=3\cdot10^{-3}\omega_0$ fixed and varying $N$ and thus $g{ _N}=\sqrt{N}\chi$. 
The vertical solid black lines correspond to the cut shown in Fig.~\ref{fig2}. 
The resonance condition is marked by dashed horizontal lines (and vertical in Fig.~\ref{fig2}). 
\label{fig33}}
\end{figure}

In Fig.~\ref{fig2}(a), we plot the GSE rates for the upper (blue curves) and lower (green curves) polariton channels versus the frequency detuning, as well as the total rate (black curves), calculating them using the JC polaritons (solid curves) and comparing them to the full bosonic model that retains the counter-rotating terms (see the SM). There is a clear inversion of the contribution to polariton creation versus detuning. 
In Fig.~\ref{fig2}(c), we plot the energy flux of such emission for $g_N=0.05\omega_0$, which shows a peak at zero detuning and shows that for $\omega_c\ll\omega_0$ the extracted energy that is associated to the emission is limited, and that there is little dependency on the detuning, both results that are in accordance to recent predictions for dissipative systems \cite{DeLiberato17}. The plots of Fig.~\ref{fig2}(a), (c) give indications for experiments based on electric current measurement. 

In a photo-detection spectroscopic experiment, the extra-cavity photonic emission is the product of two processes: First, there is the polariton scattering due to the extraction of an electron, calculated in $ \Gamma_\text{GSE}$, and which we have shown to be dependent on the $\ket{G_N}\rightarrow|\pm_{N-1}\rangle$ channel. Then there is a second relaxation process that involves the emission of a photon, 
$|\pm_{N-1}\rangle\rightarrow|G_{N-1}\rangle$, occurring with probability $|\alpha_\mathrm{ph}^{\pm}|^2$, proportional to the Hopfield coefficients associated to light (see SM for details).
Thus the multielectron GSE scattering first creates real polaritons in the cavity, which will then escape by emitting photons at their own eigenfrequencies. The spectrum of the system is thus made of two Lorentzian peaks at the polariton frequencies $\omega^-$ and $\omega^+$. 
The effective total photon emission rate, $\Gamma_\mathrm{tot}=\Gamma^{+}_\mathrm{tot}+\Gamma^{-}_\mathrm{tot}$, needs to take into account also the efficiency of this conversion, which is determined by the cavity characteristic rate, $\Gamma_\text{cav}$, and by the rate of conversion of the bright polaritons into dark polaritons, $\Gamma_\text{dark}^\pm$. For a leaky cavity, $\Gamma_\text{cav}\gg\Gamma_\text{dark}^\pm$, we have  
\begin{eqnarray}
\Gamma^{\pm}_\mathrm{tot}&=&|\alpha_\mathrm{ph}^{\pm}|^2\frac{\Gamma_\mathrm{em}^\pm\Gamma_\mathrm{cav}}{\Gamma_\text{dark}^\mathrm{\pm}+\Gamma_\mathrm{cav}}\simeq|\alpha_\mathrm{ph}^{\pm}|^2\Gamma_\mathrm{em}^\pm.
\label{totgammapm}
\end{eqnarray}
In Fig.~\ref{fig2}(b),(d), we plot such total photon emission rates and fluxes, which show that the decrease in photon collection is very modest, with respect to the electric signal measurement [Fig.~\ref{fig2}(a),{ (c)}], by which they are normalized. 
In Fig.~\ref{fig33}, we show the polariton emission rate and flux versus cavity-matter frequency detuning and of the coupling, keeping fixed the single particle coupling constant, $\chi$, which characterizes solid state structures with flat bands, so that we can span a wide range of effective light-matter coupling values, up to $g_N=0.1\omega_0$. Since $g_N=\sqrt{N}\chi$, moving rightwards in the parameter space can be achieved simply by increasing the number of emitters in the system without requiring the light-matter coupling of the microscopic model to be ultrastrong. The contour plot relative to the emission rate ($\Gamma^\pm_\text{em}$), Fig.~\ref{fig33}(a), shows an asymmetry in the detuning, favoring, at fixed $g{ _N}$, the lower polariton emission. The flux ($\omega^\pm\Gamma^\pm_\text{em}$), Fig.~\ref{fig33}(b), shows that the small emission frequency of the lower polariton curbs down the asymmetry, consistently with previous predictions \cite{DeLiberato17}.

\paragraph*{Realizations.---}
The interplay of collective photonic effects in the presence of local dissipation and transport has been recently studied in several many-body fermionic systems \cite{Keeling14,Chen14,Piazza14,Manzoni17,Shammah18,Kiffner18,Hagenmuller18,Mazza18,Zhang18b,Dinu18}.
Doped semiconductor quantum wells offer a many-body platform in which transport and ground-state properties of cavity QED can be investigated 
\cite{Khalifa08,Tsintzos08,Sapienza08,DeLiberato08,DeLiberato09,Gunter09,Lodden11,Geiser12b,Scalari12,Scalari13,DeLiberato13b,Scalari14,Paravicini18}. 

Intersubband transitions in the conduction band of these systems (the first devices to reach the ultrastrong coupling regime \cite{Gunter09}) allow to dynamically control the electron density with external fields \cite{Lolli15,Astafiev07,Stockklauser15,Liu14,Khalifa08,Tsintzos08,Sapienza08,Lodden11,Jouy10,Geiser12b}. Thus the multielectron GSE would be an effect relatively easy to explore in experiments compared to other features arising from vacuum fluctuations, such as the non-adiabatic modulation of the coupling strength.

Other candidate systems are superconducting circuits \cite{Devoret00,You05,You05b,You07,Devoret07,Rodrigues07,Hauss08,Ashhab09,You11,Gu17} and hybrid solid-state architectures
\cite{Delbecq11,Petersson12,Frey12,Frey12b,Delbecq13,Xiang13,Toida13,Wallraff13,KenaCohen13,Gubbin14,Schiro14,Viennot14,Angerer18,Stammeier18,Scarlino17,Scarlino18}, 
especially quantum-dot based systems \cite{Mu92,McKeever03,Herrmann07,Lambert09,Leturcq09,Delbecq11,Delbecq13,Lambert13b,Bergenfeldt13,Viennot14,Liu15,Deng15,Lambert15,Samkharadze16,Mi17,Landig18,Mi18,Samkharadze18}. 

\paragraph*{Conclusions.---} In conclusion, we have described a novel mechanism for light emission controlled by an electric current, occurring from the ground state of a many-body dissipative light-matter system in the regime of ultrastrong light-matter coupling.
\paragraph*{Acknowledgements.---}
\label{sec:ack}
M.C. acknowledges support from NSAF No. U1730449.
N.L. and F.N. acknowledge support from the RIKEN-AIST Challenge Research Fund, and the John Templeton Foundation.
S.D.L. acknowledges support from a Royal Society research fellowship and from the Philip Leverhulme prize of the Leverhulme Trust.
N.L. acknowledges partial support from Japan Science and Technology Agency (JST) (JST PRESTO Grant No. JPMJPR18GC). 
F.N. is partly supported by the MURI Center for Dynamic Magneto-Optics via the Air Force Office of Scientific Research (AFOSR) (Grant No. FA9550-14-1-0040), Army Research Office (ARO) (Grant No. W911NF-18-1-0358), Asian Office of Aerospace Research and Development (AOARD) (Grant No. FA2386-18-1-4045), Japan Science and Technology Agency (JST) (the Q-LEAP program, the ImPACT program and CREST Grant No. JPMJCR1676), Japan Society for the Promotion of Science (JSPS) (JSPS-RFBR Grant No. 17-52-50023, JSPS-FWO Grant No. VS.059.18N). 
\bibliography{references9b}

\clearpage
\onecolumngrid

\setcounter{equation}{0}
\setcounter{figure}{0}
\setcounter{table}{0}
\setcounter{page}{1}
\makeatletter
\renewcommand{\theequation}{S\arabic{equation}}
\renewcommand{\thefigure}{S\arabic{figure}}
\renewcommand{\bibnumfmt}[1]{[S#1]}
\renewcommand{\citenumfont}[1]{S#1}
\renewcommand{\thepage}{S\arabic{page}}
\begin{center}
\textbf{\large Supplemental Material to ``Multielectron Ground State Electroluminescence"}
\vspace*{0.2cm}

Mauro Cirio$^{1,2,*}$, Nathan Shammah$^{2,*}$, Neill Lambert$^{2}$, Simone De Liberato$^{3}$, and Franco Nori$^{2,4}$
\vspace*{0.2cm}
\emph{\small \\
$^{1}$Graduate School of China Academy of Engineering Physics, Beijing 100193, China\\
$^{2}$Theoretical Quantum Physics Laboratory, RIKEN Cluster for Pioneering Research, Wako-shi, Saitama 351-0198, Japan\\
$^{3}$School of Physics and Astronomy, University of Southampton, Southampton SO17 1BJ, United Kingdom\\
$^{4}$Department of Physics, University of Michigan, Ann Arbor, MI 48109-1040, USA\\}
\vspace*{0.2cm}
\end{center}
 
\tableofcontents
%\onecolumngrid
 \section*{Overview of the Supplemental Material}

The Supplemental Material (SM), which comprises several sections, can be divided in three main parts:

The first part is given in Sec.~\ref{sec:env},~\ref{sec:dickemod} where, firstly, we perform the mathematical passages needed to properly \emph{trace out the environmental degrees of freedom} in second quantization; secondly, we recast the fermionic Hamiltonian in terms of collective spins exploiting the symmetries of the light-matter interaction. We derive the Dicke model by reabsorbing the diamagnetic term in the photonic terms.
\blfootnote{*These two authors contributed equally to this work.}
The second part, comprising Secs.~\ref{jcp}-\ref{app:A}, illustrates the details of the \emph{bosonic model}, in which a Holstein-Primakoff approximation is performed to treat the matter-like excitations.
In Sec.~\ref{jcp} we derive the polariton eigenstates using a perturbation expansion beyond the rotating-wave approximation (RWA) on the bosonic Jaynes-Cummings (JC) part of the Hamiltonian.
In Sec.~\ref{app:A} instead we provide an alternative route, explicitly deriving the polariton eigenstates beyond the perturbation theory used in the main text, by diagonalizing, in Sec.~\ref{app:A1}, the full bosonic Hamiltonian, which contains the counter-rotating wave terms, and then deriving the transition matrix elements in terms of such states, Sec.~\ref{sec:rates}, 
showing qualitative and quantitative agreement with the perturbation theory. In Sec.~\ref{phh}, we also estimate the effective photonic emission from the cavity after the polariton scattering rates. 

The third part of the SM instead, comprising Sec.~\ref{sec:appD}, develops the \emph{fermionic model} [\Eq{eq:full} of the main text], retaining the full nonlinearities for the matter-like excitations.
Following this route, we will arrive to an expression for the single-polariton processes involved in the many-body ground state electroluminescence that can be compared to the one obtained for the bosonic models. Moreover, in this formalism, we can capture more confidently also the scattering processes involving double-polariton states.
However, in this framework, we will need to develop a more refined second-quantization theory capable of grasping the microscopic processes involved in this effect, to leading order in perturbation theory, with light-matter states comprising both fermionic and bosonic degrees of freedom.
In order to do so, we derive a model for electron addition and subtraction, which will lead us to map the electron current effects in terms of Dicke states with non-fixed particle number, beyond a RWA analysis.

In particular, in Sec.~\ref{sec:appD}, using the Dicke state formalism, we develop perturbative fermionic eigenstates in terms of the RWA Hamiltonian of the light-matter system.
In Sec.~\ref{app:RWA}, we highlight the properties of the Tavis-Cummings model eigenstates, used to derive the perturbed basis of the Dicke model.
In Sec.~\ref{sec:appF1}, we introduce the difference between the microscopic and macroscopic state formalism and derive the corresponding master equations.
In Sec.~\ref{sec:appF}, we calculate the general transition matrix elements induced by the interaction Hamiltonian relative to the electronic environment.
In Sec.~\ref{sec:appG}, we unravel the compact but cumbersome formal expression derived in Sec.~\ref{sec:appF} and specialize it to the case of ground state electroluminescence, calculating both the single-polariton and double-polariton transition rates. Finally, in Sec.~\ref{sec:appGl}, we compare the fermionic results with the bosonic calculations, finding excellent agreement.

\section{Tracing out the environment}
\label{sec:env}
We begin by considering \Eq{eq:full} of the main text, 
\begin{eqnarray}
\label{eq:full:supp}
H&=&\omega_{c}a^\dagger a+\displaystyle\sum_n\left(\omega_1 c_{1,n}^\dagger c_{1,n} + \omega_2 c_{2,n}^\dagger c_{2,n}\right)+D(a+a^\dagger)^2
+\chi(a+a^\dagger) \displaystyle\sum_{n}(c_{2,n}^\dagger c_{1,n}+c_{1,n}^\dagger c_{2,n}),
\end{eqnarray}
which describes light-matter interaction in the model system under study. 
We are interested in studying the effects of three environments on this model: a left {($L$)} and right {($R$)} electronic reservoir, and the extra-cavity electromagnetic modes. The aim of this section is to compute the {transition rates} among eigenstates of the system induced by a generic Hamiltonian $H^I=H^I_{\text{el}}+H^I_{\text{cav}}$ representing the physical interaction with the electronic and bosonic environmental degrees of freedom. 
By considering $H^I$ as a perturbation over the coherent dynamics, the electron scattering rates can be computed by using the Fermi golden rule
\begin{equation}
\label{eq:FGR}
\Gamma^{\alpha\rightarrow \beta}_{\text{el}}=2\pi\sum_i\rho^\text{FD}_i\sum_f|\bra{\beta,f}H^I_{\text{el}}\ket{\alpha,i}|^2\delta(\tilde{\Delta}),
\end{equation}
where $\alpha$ ($i$) and $\beta$ ($f$) are the initial and final states for the system (environment).
The fermionic part of the environment follows a Fermi-Dirac distribution $\rho^{\text{FD}}$, diagonal in the basis $\ket{i}$. This distribution depends on macroscopic parameters characterizing the reservoir, such as the temperature and chemical potential. The energy conservation is imposed by the delta function with argument {$\tilde{\Delta}=\omega_f-\omega_i+\Delta_{\alpha\beta}$, with $\Delta_{\alpha\beta}=\omega_\beta-\omega_\alpha$, where $\omega_\beta$ and $\omega_\alpha$ are the final and initial frequencies of the system.}
Specifically, we model the interaction with the electronic reservoirs as $H^I_\text{el}=H^I_{L}+H^I_{R}$, where
\begin{equation}
\label{eq:HLRapp}
H^I_{L}=\lambda\sum_{n,\zeta}[(c_{1,n}+c_{2,n})c^\dagger_{L;n,\zeta}+(c^\dagger_{1,n}+c^\dagger_{2,n})c_{L;n,\zeta}],
\end{equation}
and similarly for $H^I_{R}$, thus assuming that the energy scale, $\lambda$, is equal for the two fermionic reservoirs.
The operators {$c_{L(R);n,\zeta}$} label the annihilation operators for a fermion associated with a degree of freedom $n$ and $\zeta$ in the left (right) reservoir.
The label $\zeta$ represents a continuum of properties, so we replace $\sum_{\zeta}\rightarrow\int d\zeta\;\nu(\zeta)$ with a density of states $\nu(\zeta)$.
By introducing a generic dispersion relation of the kind $E=f(\zeta)$ we can further write $\sum_{\zeta}\rightarrow\int d\zeta\;\nu(\zeta)=\int dE\;\nu(E)$, so that $\nu(\zeta)=\frac{d\zeta}{dE}\nu(E)$ and we obtain
\begin{equation}
\label{eq:Gamma_Gamma_n}
\Gamma^{\alpha\rightarrow \beta}_\text{el}=\sum_n\Gamma^{\alpha\rightarrow \beta}_{\text{el},n}
\end{equation}
with
\begin{eqnarray}
\nonumber
\Gamma^{\alpha\rightarrow \beta}_{\text{el},n}&=&2\pi\lambda^2\sum_i\rho^\text{FD}_i\left[\int dE\;\nu(E)\left[{1-n_i(E)}\right]\delta(\tilde{\Delta})|\bra{\beta}(c_{1,n}+c_{2,n})\ket{\alpha}|^2\right.\\
&&\left.+\int dE\;\nu(E){n_i(E)}\delta(\tilde{\Delta})|\bra{\beta}(c^\dagger_{1,n}+c^\dagger_{2,n})\ket{\alpha}|^2\right]=\Gamma^{\alpha\rightarrow \beta}_{\text{out},n}+\Gamma^{\alpha\rightarrow \beta}_{\text{in},n}\;,
\label{eq:partPart}
\end{eqnarray}
where $n_i(E)$ counts the number of electrons with energy $E$ in the state $\ket{i}$ of the reservoir according to the Fermi-Dirac distribution $\rho^\text{FD}_i$.
The rate $\Gamma^{\alpha\rightarrow \beta}_{\text{out},n}$ ($\Gamma^{\alpha\rightarrow \beta}_{\text{in},n}$) appearing in the r.h.s.~of \Eq{eq:partPart} accounts for processes in which the environment has one more (less) electron associated with some quantum number $\zeta$, hence the definition of in/out rates:
\begin{subequations}
\begin{eqnarray}
\Gamma^{\alpha\rightarrow \beta}_{\text{out}}&=&\sum_n \Gamma^{\alpha\rightarrow \beta}_{\text{out},n}
=\sum_n2\pi\lambda^2
\left[1-\bar{n}_\text{FD}(-\Delta_{\alpha\beta})\right]\nu(-\Delta_{\alpha\beta})|\bra{\beta}(c_{1,n}+c_{2,n})\ket{\alpha}|^2,\\
\Gamma^{\alpha\rightarrow \beta}_{\text{in}}&=&\sum_n \Gamma^{\alpha\rightarrow \beta}_{\text{in},n}
=\sum_n2\pi\lambda^2
\bar{n}_\text{FD}(\Delta_{\alpha\beta})\nu(\Delta_{\alpha\beta})|\bra{\beta}(c^\dagger_{1,n}+c^\dagger_{2,n})\ket{\alpha}|^2,
\end{eqnarray}
\label{eq:rates}
\end{subequations}
where $\bar{n}_\text{FD}(E)=\sum_i\rho^\text{FD}_in_i(E)$ is the average electron number in the left or right electronic reservoir with energy $E$.
In Eqs.~(\ref{eq:Gamma_Gamma_n}),(\ref{eq:partPart}),(\ref{eq:rates}), the rates can be specified to the left and right reservoir, something we omitted above for simplicity, but which will hereafter be labelled for clarity in the \emph{electron-current transition rates}
\begin{equation}
\label{eq:rates_0app}
\Gamma^{\alpha\rightarrow \beta}_{L(R)}=\Gamma^{\alpha\rightarrow \beta}_{L(R),\text{out}}+\Gamma^{\alpha\rightarrow \beta}_{L(R),\text{in}}.
\end{equation}
If the electronic reservoirs are at zero temperature, the Fermi-Dirac distribution is only a function of the chemical potential, $\mu_{L}$ and $\mu_{R}$ respectively, such that we obtain
a very compact expression relating the transition matrix elements to the ``in" and ``out" electron-current transition rates
\begin{equation}
\label{eq:ratesT0}
\begin{array}{lll}
\Gamma^{\alpha\rightarrow \beta}_{L(R),\text{out}}&=&\Gamma_\text{el}~{\theta(-\mu_{L(R)}-\Delta_{\alpha\beta}})\sum_n|M_{\alpha\beta}^n|^2,\\
\Gamma^{\alpha\rightarrow {\beta}}_{L(R),\text{in}}&=&\Gamma_\text{el}~\theta(\mu_{L(R)}-\Delta_{\alpha\beta})\sum_n|M_{\beta\alpha}^n|^2,\\
\end{array}
\end{equation}
where $\Gamma_\text{el}=2\pi\lambda^2\nu$ and $\theta(x)$ is the Heaviside function.
At zero temperature it is possible to define a frequency-independent density of states $\nu$ for the two electronic reservoirs.
We recall that the \emph{electron-current transition matrix element}, introduced in \Eq{eq:M} in the main text, reads
\begin{equation}
\label{eq:Mapp}
M_{\alpha\beta}^n=\bra{\beta}(c_{1,n}+c_{2,n})\ket{\alpha},
\end{equation}
where we consider that an electron in the upper or lower band can be lost, given the fact that, as we will show in the next Section, the light-matter ground state of the Hamiltonian in \Eq{eq:full}, also \Eq{eq:full:supp}, contains a superposition of such populations.

Similarly {to \Eq{eq:HLRapp}}, the interaction {of the cavity field} with the extra-cavity electromagnetic modes is described by
\begin{equation}
\label{eq:Hcav}
H^I_\text{cav}=\lambda_\text{cav} \sum_{\zeta}(a+a^\dagger)(a_\zeta+a_{\zeta}^\dagger),
\end{equation}
where $\lambda_\text{cav}$ is the photon-photon coupling strength and
$a_\xi$ is the annihilation operator for the $\xi$th {extra-cavity} mode.
By analyzing this formula, in analogy to what done for the electronic environment, we can derive the \emph{cavity-photon emission rate},
\begin{eqnarray}
\Gamma^{\alpha\rightarrow \beta}_{\text{cav}}&=&2\pi\sum_i\rho^{\text{BE}}_i\sum_f|\bra{\beta,f}H^I_{\text{cav}}\ket{\alpha,i}|^2\delta(\tilde{\Delta})
%\nonumber\\&=&
=2\pi\lambda_\text{cav}^2[\bar{n}_\text{BE}(\Delta_{\alpha\beta})\nu_{\text{EM}}(\Delta_{\alpha\beta})
%\nonumber\\&&
+\left[1+\bar{n}_\text{BE}(-\Delta_{\alpha\beta})\right]\nu_\text{EM}(-\Delta_{\alpha\beta})]{M^\text{cav}_{\alpha\beta}}^2,\nonumber\\
\label{eq:GammaCavT0}
\end{eqnarray}
where $\rho^{\text{BE}}_i$ is the Bose-Einstein distribution for photons in the state $\ket{i}$ and $\bar{n}_\text{BE}$ counts the average number of photons in the electromagnetic reservoir. Defining the \emph{cavity-photon transition matrix elements} as
\begin{equation}
\label{eq:M2}
M^\text{cav}_{\alpha\beta}=\bra{\beta}(a+a^\dagger)\ket{\alpha}.
\end{equation}
At zero temperature and for a frequency-independent photonic density of states, $\nu_{\text{EM}}(\omega)\simeq\nu_{\text{EM}}$, we {obtain}
\begin{equation}
\label{eq:GammaCavT0}
\begin{array}{lll}
\Gamma^{\alpha\rightarrow \beta}_\text{cav}&=&\Gamma_\text{cav}{M^\text{cav}_{\alpha\beta}}^2,
\end{array}
\end{equation}
where $\Gamma_\text{cav}=2\pi\lambda_\text{cav}^2\nu_{\text{EM}}$ {is the characteristic emission rate of the photonic cavity}.

\section{Dicke model}
\label{sec:dickemod}

We can rewrite \Eq{eq:full:supp} in terms of spin angular momentum operators. We define $S^\pm=\sum_n S_{n}^\pm$, with $S_{n}^+= c_{2,n}^\dagger c_{1,n}$ and thus
\begin{eqnarray}
2S^3&=&2\sum_n S_{n}^3=\sum_n\sigma_{z,n}=\sum_n \lbrack S_{n}^+,S_{n}^- \rbrack=\sum_n \left(c_{2,n}^\dagger c_{2,n}-c_{1,n}^\dagger c_{1,n}\right).
\end{eqnarray}
We immediately find that the total angular momentum, $\mathbf{S}^2$, is a symmetry of the model. We obtain
\begin{eqnarray}
H&=&\omega_{c}a^\dagger a+\omega_0 S^3 +\chi(a+a^\dagger)(S^{-}+S^+)+D(a+a^\dagger)^2
+\tilde{E}_0(N,N_2),
\end{eqnarray}
where we shifted the energy to absorb a term 
\begin{eqnarray}
E_0(N,N_2)=\omega_1N+2\omega_1N_2+\omega_0(j_N+N_2)
\end{eqnarray}
with $j_N=N/2$. 
We performed a fermion-to-spin transformation which, with respect to \Eq{eq:full:supp}, involves no approximations. 
In order to absorb the diamagnetic term, we rewrite the bosonic operator for the light field as a displaced operator, 
\begin{eqnarray}
\tilde{a}&=&\cosh(\lambda)a+\sinh(\lambda)a^\dagger, 
\end{eqnarray}
which allows us to rewrite the Hamiltonian of Eq.~(1) in the main text, repeated as 
\begin{eqnarray}
H_{N,N_2}&=&\tilde{\omega}_{c}\tilde{a}^\dagger \tilde{a}+\omega_0 S^3 +\tilde{\chi}(\tilde{a}+\tilde{a}^\dagger)(S^{-}+S^+)
%\nonumber\\&&
+\tilde{E}_0(N,N_2),
\end{eqnarray}
where 
\begin{eqnarray}
\tilde{\omega}_c&=&\omega_{c}e^{2\lambda},\\
 \tilde{\chi}&=&\chi e^{-\lambda},\\
 \tilde E_0&=&\frac{\omega_{c}}{2}\left( e^{-2\lambda}-1\right)+E_0,\\
 \lambda&=&\frac{1}{2}\text{arctanh}[{D}/{(\omega_c+2D})].
 \end{eqnarray}

 Renaming $\tilde{a}\rightarrow a$, $\tilde{\omega}_c\rightarrow {\omega}_c$, and $\tilde{\chi}\rightarrow \chi$
as implicitly assumed elsewhere in the text, and reabsorbing the constant term $\tilde{E}_0$, we obtain Eq.~(2) of the main text, the Hamiltonian of the Dicke model,
\begin{eqnarray}
\label{eq:simplified:sup}
H&=&\omega_{c}a^\dagger a+\omega_0 S^3 +\chi(a+a^\dagger)(S^{-}+S^+).
\end{eqnarray}

\section{Bosonic model: Perturbative theory of the bosonic model}
\label{jcp}
In the main text, the polariton modes of the JC model are defined in terms of the \emph{rotating-wave part} of the bosonic Hamiltonian, \Eq{eq:Hbos}. Here we illustrate explicitly the passages involved to derive the polariton eigenstates involved in the transition processes, developing them perturbatively onto the bosonic Jaynes-Cummings polariton eigenstates.
\subsection{Holstein-Primakoff transformation}
\label{sec:hp}

The introduction of bosonic operators for the matter-excitations,
$S_+=\sqrt{N}b^\dagger+O(|{b^\dagger b}/{\sqrt{N}}|)$, and $S_z= b^\dagger b-j_N$ 
%eq
allow us to rewrite \Eq{eq:simplified:sup} as
\begin{eqnarray}
\label{eq:fullboso}
H_\text{bos}=\omega_c a^\dagger a+\omega_0 b^\dagger b+ g_N (a^\dagger+  a)(b + b^\dagger ),
\end{eqnarray}
by performing an approximation that is valid in the dilute regime, $|{b^\dagger b}/{\sqrt{N}}|\ll 1$. 
In the following subsections we provide the explicit derivation of the perturbed polariton eigenstates on the RWA part of this bosonic Hamiltonian.
\subsection{Perturbative eigenstates of the bosonic model}
\label{sec:jcm}
In the main text, we restricted ourselves to diagonalize the Jaynes-Cummings part of the bosonic Hamiltonian in Eq.~(\ref{eq:Hbos}), rewritten above as \Eq{eq:fullboso}, i.e.,
\begin{equation}
H_\text{JC}=\omega_c a^\dagger a+\omega_0 b^\dagger b+ g_N (a^\dagger b + b^\dagger a),
\end{equation}
whose kernel reads
\begin{equation}
\bar{H}=
\left(\begin{array}{cccc}
\omega_c&0&g_N&0\\
0&-\omega_c&0&-g_N\\
g_N&0&\omega_0&0\\
0&-g_N&0&-\omega_0
\end{array}\right).
\end{equation}
and its polariton modes are given as
\begin{equation}
\label{jcpol}
p_\pm^\dagger = \alpha^\pm_a a^\dagger + \alpha^\pm_b b^\dagger,
\end{equation}
with
\begin{equation}
\begin{array}{lll}
\alpha_a^\pm&=&\pm\displaystyle\left(\frac{\sqrt{4+x^2}\mp x}{2\sqrt{4+x^2}}\right)^\frac{1}{2},\\
\alpha_b^\pm&=&\displaystyle\left(\frac{2}{\sqrt{4+x^2}\left(\sqrt{4+x^2}\mp x\right)}\right)^\frac{1}{2},
\end{array}
\label{aaa}
\end{equation}
where 
\begin{eqnarray}
x&=&\Delta/g_N
\end{eqnarray}
with 
\begin{eqnarray}
 \Delta&=&\omega_0-\omega_c.
 \end{eqnarray}
 Note that through \Eq{aaa}, it becomes explicit the dependence on $N$ of the polariton modes, $p_\pm=p_\pm^N$, which is implicit in \Eq{jcpol}. Hereafter, for simplicity, we will omit this explicit dependence besides for the expressions in which it will be clearer to highlight it.

In terms of these modes the {bosonic} Jaynes-Cummings Hamiltonian reads
\begin{eqnarray}
H_\text{JC}&=&\omega_+ p_+^\dagger p_+ + \omega_- p_-^\dagger p_-,
\end{eqnarray}
with 
\begin{eqnarray}
\omega_\pm &=& \frac 1 2 \left(\omega_0+\omega_c\pm\sqrt{4 g_N^2+\Delta^2}\right).
\end{eqnarray}
As a consequence, the full Hamiltonian now reads
\begin{equation}
H=H_\text{JC}+V,
\end{equation}
with
\begin{eqnarray}
V&=&g_N\left[ -\alpha^+_a\alpha^+_b p_-^\dagger p_-^\dagger - \alpha^-_a\alpha^-_b p_+^\dagger p_+^\dagger
%\right.\nonumber\\&& \left.
 +(\alpha^+_b\alpha^-_a+\alpha^-_b\alpha^+_a) p_-^\dagger p_+^\dagger\right]+\text{H.c.}
\end{eqnarray}
where we used the identity $\alpha^-_a\alpha^+_b-\alpha^-_b\alpha^+_a=-1$ to write the operators $a$ and $b$ in terms of the Jaynes-Cummings polaritons {of \Eq{jcpol}}. We can then define the eigenstates of the {bosonic} Jaynes-Cummings Hamiltonian as
\begin{subequations}
\label{states0}
\begin{eqnarray}
\ket{G^{(0)}_N}&=&\bigotimes_{n\in N}c^\dagger_{1,n}\ket{0_\text{el}}\ket{0_\text{ph}},\\
\ket{\pm^{(0)}_N}&=&p^\dagger_\pm \ket{G^{(0)}_N},\\
\ket{\pm\pm^{(0)}_N}&=&\displaystyle\frac{p^\dagger_\pm p^\dagger_\pm}{\sqrt{2}} \ket{G^{(0)}_N},\\
\ket{+-^{(0)}_N}&=&p^\dagger_+ p^\dagger_- \ket{G^{(0)}_N},
\end{eqnarray}
\end{subequations}
where $\ket{0_\text{el}}$ and $\ket{0_\text{ph}}$ are the electron and photonic vacuum, respectively.
By using first-order perturbation theory in the potential $V$ we then find that
\begin{subequations}
\begin{eqnarray}
\ket{G_N}&=&\ket{G^{(0)}_N}-\beta_{++}\ket{++^{(0)}_N}%\nonumber\\&&
-\beta_{--}\ket{--^{(0)}_N} -\beta_{+-}\ket{+-^{(0)}_N},\\
\ket{\pm_N}&=&\ket{\pm^{(0)}_N}+\cdots,\\
\ket{\pm\pm_N}&=&\ket{\pm\pm^{(0)}_N}+\beta_{\pm\pm}\ket{G^{(0)}_N}+\cdots,\\
\ket{+-_N}&=&\ket{+-^{(0)}_N}+\beta_{+-}\ket{G^{(0)}_N}+\cdots,
\end{eqnarray}
\label{eqst}
\end{subequations}
with
\begin{subequations}
\label{eqbeta}
\begin{eqnarray}
\beta_{++}&=&\bra{++^{(0)}_N} V\ket{G^{(0)}_N}=-\displaystyle\sqrt{2}\frac{g_N}{2\omega_+}\alpha^-_a\alpha^-_b,\\
\beta_{--}&=&\bra{--^{(0)}_N} V\ket{G^{(0)}_N}=-\displaystyle\sqrt{2}\frac{g_N}{2\omega_-}\alpha^+_a\alpha^+_b,\\
\beta_{+-}&=&\bra{+-^{(0)}_N} V\ket{G^{(0)}_N}=\displaystyle\frac{g_N}{\omega_++\omega_-}(\alpha^+_b\alpha^-_a + \alpha^-_b\alpha^+_a),%\nonumber\\
\end{eqnarray}
\end{subequations}
which are the results used in the main text. Note that also the coefficients $\beta_{xy}$ have a dependence on $N$.
\subsection{Commutation relations between Jaynes-Cummings polaritons and electrons}
\label{sec:appB1}
We wish to calculate the commutators 
\begin{eqnarray}
C_{1,n} &=& [b^\dagger,c_{1,n}], \;\;\;\;\;\; C_{2,n} = [b^\dagger,c_{2,n}], 
\end{eqnarray}
in order to calculate the electron-current transition matrix elements within the RWA bosonic model. 
Using the approximate Holstein-Primakoff transformation
\begin{equation}
b^\dagger=\frac{S_+}{\sqrt{N}}=\sum_{\bar{n}} \frac{c^\dagger_{2,\bar{n}}c_{1,\bar{n}}}{\sqrt{N}},
\end{equation}
we obtain
\begin{equation}
\begin{array}{lll}
\sqrt{N}C_{1,n} &=&\sum_{\bar{n}} c^\dagger_{2,\bar{n}}c_{1,\bar{n}} c_{1,{n}}- c_{1,{n}} c^\dagger_{2,\bar{n}}c_{1,\bar{n}},\\
\sqrt{N}C_{2,n}&=&\sum_{\bar{n}} c^\dagger_{2,\bar{n}}c_{1,\bar{n}} c_{2,{n}} - c_{2,{n}} c^\dagger_{2,\bar{n}}c_{1,\bar{n}}.
\end{array}
\end{equation}
We now note that, if $\bar{n}\neq{n}$ the fermionic anticommutation rules give zero. As a consequence, the sum only contributes for $\bar{n}={n}$ giving
\begin{equation}
\begin{array}{lll}
\sqrt{N}C_{1,n} &=& c^\dagger_{2,{n}}c_{1,{n}} c_{1,{n}}- c_{1,{n}} c^\dagger_{2,{n}}c_{1,{n}}=0\\
\sqrt{N}C_{2,n}&=& c^\dagger_{2,{n}}c_{1,{n}} c_{2,{n}} - c_{2,{n}} c^\dagger_{2,{n}}c_{1,{n}}=-c_{1,{n}},
\end{array}
\end{equation}
where we used again the fermionic anticommutation rules.\\
Similarly, it is possible to calculate the analogous commutators involving $b$, resulting in the following commutators
\begin{equation}
\label{eq:comm}
\begin{array}{lll}
{}[b,c_{1,n}]&=&-\displaystyle\frac{1}{\sqrt{N}}c_{2,n},\\
{}[b,c_{2,n}]&=&0,\\
{}[b^\dagger,c_{1,n}]&=&0,\\
{}[b^\dagger,c_{2,n}]&=&-\displaystyle\frac{1}{\sqrt{N}}c_{1,n},
\end{array}
\end{equation}
which provides a prescription for the actual calculation of the electron-current transition matrix elements, performed in the following subsection.
\subsection{Transition matrix elements in the bosonic model using Jaynes-Cummings polaritons}
\label{sec:appB}
We now calculate the transition matrix elements between the ground state and the first excited states.
The transition to single-polariton states, $M_{G\pm}^n$, using the results of the perturbation theory, Eqs.~(\ref{states0}),(\ref{eqst}),(\ref{eqbeta}), gives, to first order in the perturbative parameters $\beta_{ij}$ given in \Eq{eqst}
\begin{eqnarray}
M_{G\pm}^n&=&\bra{\pm_{N-1}}(c_{1,n}+c_{2,n})\ket{G_N}
%\nonumber\\&=&
=\bra{G^{(0)}_{N-1}}p_\pm (c_{1,n}+c_{2,n})\ket{G_N}\nonumber\\
&=&\bra{G^{(0)}_{N-1}}p_\pm (c_{1,n}+c_{2,n})\ket{G^{(0)}_N}
%\nonumber\\&&
-\displaystyle\frac{\beta_{++}}{\sqrt{2}}\bra{G^{(0)}_{N-1}}p_\pm (c_{1,n}+c_{2,n})p^\dagger_+p^\dagger_+\ket{G^{(0)}_N}
\nonumber\\&&
-\displaystyle\frac{\beta_{--}}{\sqrt{2}}\bra{G^{(0)}_{N-1}}p_\pm (c_{1,n}+c_{2,n})p^\dagger_-p^\dagger_-\ket{G^{(0)}_N}
%\nonumber\\&&
-\beta_{+-}\bra{G^{(0)}_{N-1}}p_\pm (c_{1,n}+c_{2,n})p^\dagger_+p^\dagger_-\ket{G^{(0)}_N}.%\nonumber\\
\label{mgpmpert}
\end{eqnarray}
The first term on the r.h.s.~of \Eq{mgpmpert} is $0$, as $c_{2,n}$ is annihilated on the Fermi sea at zero temperature and $c_{1,n}$ describes the Fermi sea without one electron, i.e.,
\begin{subequations}
\begin{eqnarray}
c_{2,n}\ket{G^{(0)}_{N}}&=&0\\
c_{1,n}\ket{G^{(0)}_{N}}&=&\ket{G^{(0)}_{N-1}}.
\end{eqnarray}
\label{c1c2}
\end{subequations}
In \Eq{mgpmpert}, the average of a polariton creation operator on the ground state is zero.
To unravel the other terms in \Eq{mgpmpert} it is useful to calculate
  \begin{eqnarray}
{}[c_{1,n}+c_{2,n},p^\dagger_\pm]=\alpha_b^{\pm}[c_{1,n}+c_{2,n},b^\dagger]=\alpha_b^{\pm}\frac{c_{1,n}}{\sqrt{N}}.
  \label{comm}
 \end{eqnarray}
Equation~(\ref{comm}) implies that the second term on the r.h.s.~of \Eq{mgpmpert}, proportional to $\beta_{++}$, can be written as
 \begin{eqnarray}
\bra{G^{(0)}_{N-1}}p_\pm (c_{1,n}+c_{2,n})p^\dagger_+p^\dagger_+\ket{G^{(0)}_N}&=&
\alpha_b^{\pm}\frac{1}{\sqrt{N}}\bra{G^{(0)}_{N-1}}p_\pm c_{1,n}p^\dagger_+\ket{G^{(0)}_N}
+\bra{G^{(0)}_{N-1}}p_\pm p^\dagger_+(c_{1,n}+c_{2,n}) p^\dagger_+\ket{G^{(0)}_N}\nonumber\\
&=&2\alpha_b^{\pm}\frac{1}{\sqrt{N}}\bra{G^{(0)}_{N-1}}p_\pm p^\dagger_+c_{1,n}\ket{G^{(0)}_N}
+\bra{G^{(0)}_{N-1}}p_\pm p^\dagger_+ p^\dagger_+(c_{1,n}+c_{2,n})\ket{G^{(0)}_N}\nonumber\\
&=&2\alpha_b^{\pm}\frac{1}{\sqrt{N}}\bra{G^{(0)}_{N-1}}p_\pm p^\dagger_+\ket{G^{(0)}_{N-1}}
+\bra{G^{(0)}_{N-1}}p_\pm p^\dagger_+ p^\dagger_+\ket{G^{(0)}_{N-1}}=\delta_{\pm,+}\frac{2\alpha_b^{\pm}}{\sqrt{N}},
\label{bppeq}
\end{eqnarray}
where $\delta_{\pm,+}$ is the Kr\"{o}necker delta giving 1 for $M_{G+}^n$ and 0 for $M_{G-}^n$.
As shown by the commutator in \Eq{comm}, this matrix element depends on the presence of $c_{2,n}$ in the interaction Hamiltonian, which means that this process relies on a $c_{2,n}$ electron to tunnel out of the system.

Similarly to \Eq{bppeq}, one can calculate the third and fourth term in \Eq{mgpmpert}, proportional to $\beta_{--}$ and $\beta_{+-}$, respectively.
The result is then
 \begin{eqnarray}
 M_{G\pm}^n=\frac{\sqrt{2}\beta_{\pm\pm}\alpha_b^\pm+\beta_{+-}\alpha_b^\mp}{\sqrt{N}}.
 \label{mpferm}
 \end{eqnarray}

We now consider the transition to double-polariton states, characterized by the matrix elements $M_{G\pm\pm}^n$,
 \begin{eqnarray}
 M_{G\pm\pm}^n&=&
\bra{\pm\pm_{N-1}}(c_{1,n}+c_{2,n})\ket{G_N}%\nonumber\\&=& 
=\beta^N_{\pm\pm}\bra{G^{(0)}_{N-1}} (c_{1,n}+c_{2,n})\ket{G^{(0)}_N}
%\nonumber\\ &&
 -\frac{\beta^{N-1}_{\pm\pm}}{2}\bra{G^{(0)}_{N-1}} p_+p_+(c_{1,n}+c_{2,n})p^\dagger_\pm p^\dagger_\pm\ket{G^{(0)}_N}.\nonumber\\
 \end{eqnarray}
We already see a major difference with respect to the previous case. In fact, the first element is non-zero because of the $c_{1,n}$ term.
We also immediately see that, when calculating a commutator, the second term ends up with an odd number of electrons and vanishes.
This means that the only non-zero result will be when no commutators are computed.
At that point only $c_{1,n}$ will contribute to transform the Fermi sea with $N$ electrons to a Fermi sea with $(N-1)$ electrons.
Therefore,
 \begin{equation}
  \begin{array}{lll}
 M_{G\pm\pm} ^n&=&\beta^N_{\pm\pm}-\beta^{N-1}_{\pm\pm}{\simeq}\partial_N \beta^N_{\pm\pm},
 \end{array}
 \end{equation}
 which needs to be compared with the transition matrix element calculated in the full bosonic model, \Eq{eq:Hbos}, and the same applies for the single-polariton rates, \Eq{mpferm}.
\section{Bosonic model: Full bosonic model diagonalization beyond perturbation theory}
\label{app:A}
Here we derive the \emph{full} polariton spectrum of the Hamiltonian in \Eq{eq:Hbos}, as well as its eigenstates, and use them to calculate the transition rates between the ground state and the relevant excited states.
\subsection{Polariton spectrum of the full bosonic model}
\label{app:A1}
We {reconsider \Eq{eq:Hbos} by writing it as}
\begin{equation}
\label{HHH_full}
H=\omega_c a^\dagger a+\omega_0 b^\dagger b + g_N(a+a^\dagger)(b+b^\dagger),
\end{equation}
{whose} normal modes $\hat{v}$ can be found by solving the Hopfield equation
\begin{equation}
[H,\hat{v}]=E_{v} \hat{v}.
\end{equation}
The commutator structure of this equation implies a built-in ``time reversal'' symmetry which simply reads
\begin{equation}
[H,\hat{v}^\dagger]=-E_{\pm} \hat{v}^\dagger.
\end{equation}
By writing a generic operator 
$\hat{v}$ as 
\begin{equation}
\hat{v}=c_1 a^\dagger + c_2 a+c_3 b^\dagger+c_4 b=\sum_i v_i\cdot \hat{O}_i, 
\end{equation}
where $\vec{v}\in\mathbb{R}^4$ and $\hat{\vec{O}}=\{a^\dagger,a,b^\dagger,b\}$, we obtain
\begin{equation}
\sum_i [H,\hat{O_i}]v_i=E_v\sum_i v_i \hat{O}_i.
\end{equation}
Since the set of operators in $\vec{O}$ is closed under commutations with the Hamiltonian we can define a Hamiltonian kernel $\bar{H}$, such that
\begin{equation}
{}[H,\hat{O}_i]=\sum_j \bar{H}_{ji}\hat{O}_j.
\end{equation}
This leads to $\sum_j\sum_i \bar{H}_{ji}v_i\hat{O_j}=E_v\sum_i v_i \hat{O}_i$, or, equivalently
\begin{equation}
\sum_j\sum_i \bar{H}_{ji}v_i\hat{O}_j=E_v\sum_j v_j \hat{O}_j,
\end{equation}
whose solution can be found by solving
\begin{equation}
\begin{array}{lll}
\sum_{j}\bar{H}_{ij}v_j&=&E_v v_j,
\end{array}
\end{equation}
or, equivalently,
\begin{equation}
\bar{H}\vec{v}=E_v\vec{v}.
\end{equation}
Explicitly, the matrix $\bar{H}$ can be constructed column by column, taking the commutator of $H$ with $a^\dagger$ (first column), $a$ (second column), $b^\dagger$ (third column), and $b$ (fourth column). The entry of the first (second) row corresponds to taking the component proportional to $a^\dagger$ {($a$)} of the column by column procedure. The entry of the third (fourth) row corresponds to taking the component proportional to $b^\dagger$ {($b$)} of the column by column procedure. Explicitly we have
\begin{equation}
\bar{H}=
\left(\begin{array}{cccc}
\omega_c&0&g_N&-g_N\\
0&-\omega_c&g_N&-g_N\\
g_N&-g_N&\omega_0&0\\
g_N&-g_N&0&-\omega_0
\end{array}\right).
\end{equation}
There are two eigenvectors $\vec{v}_\pm$ of the previous equation with positive energy and they correspond to the \emph{polariton modes} of the \emph{full bosonic model}
 \begin{equation}
 \label{poo}
{\tilde{p}}^\dagger_\pm=\sum_i v^\pm_i\hat{O}_i.
\end{equation}
 The normalization follows from the imposition $[{\tilde{p}}_\pm,{\tilde{p}}_\pm^\dagger]=1$ and requires
\begin{equation}
\label{eq:constr_bosons}
{v^\pm_1}^2+{v^\pm_3}^2-{v^\pm_2}^2-{v^\pm_4}^2=1.
\end{equation}
Notice that, if we used ${\tilde{p}}_\pm$ in the definition of \Eq{poo}, a similar analysis for the operators ${\tilde{p}}_\pm$ we would have imposed $v_2^2+v_4^2-v_1^2-v_3^2=1$ instead.\\
To summarize, we now write explicitly the solution of
\begin{equation}
\bar{H}\vec{v}_\pm=\lambda_\pm \vec{v}_\pm,
\end{equation}
such that $\lambda_\pm>0$. We have that the eigenenergies of the upper and lower polariton branches are then given by
\begin{equation}
\lambda_\pm=\left[\frac{\omega_0^2+\omega_c^2\pm\left(\omega_0^4+16 g_N^2\omega_0\omega_c-2\omega_0^2\omega_c^2+\omega_c^4\right)^\frac{1}{2}}{2}\right]^\frac{1}{2},
\end{equation}
and the eigenvectors are defined by
\begin{equation}
\begin{array}{lll}
v_1^\pm&=&g_N(\lambda_\pm+\omega_0)(\lambda_\pm+\omega_c)/Z_\pm,\\
v^\pm_2&=&-g_N(\lambda_\pm+\omega_0)(\lambda_\pm-\omega_c)/Z_\pm,\\
v^\pm_3&=&[2g_N^2 \omega_c+(\lambda_\pm+\omega_0)(\lambda_\pm^2-\omega_c^2)]/Z_\pm,\\
v^\pm_4&=&-2g_N^2\omega_c/Z_\pm,
\end{array}
\end{equation}
where $Z_\pm$ is set to ensure \Eq{eq:constr_bosons}. The final expression for the polaritonic modes for the full bosonic Hamiltonian of \Eq{HHH_full} is then
\begin{equation}
\begin{array}{lll}
{\tilde{p}}^\dagger_\pm&=&\sum_i v_i^\pm\hat{O}_i
%\\&=&
=v_1^\pm a^\dagger+v_2^\pm a+v_3^\pm b^\dagger+v_4^\pm b.
\end{array}
\label{eq:fullp}
\end{equation}
Note that we can also relate $\hat{\vec{p}}=\{{\tilde{p}}_+^\dagger,
\tilde{p}_+, \tilde{p}_-^\dagger, \tilde{p}_-\}^T$ to $\hat{\vec{O}}$. 
By defining
\begin{equation}
\label{eq:pmatrix}
P=
\left(\begin{array}{cccc}
v^1_+&v^2_+&v^3_+&v^4_+\\
v^2_+&v^1_+&v^4_+&v^3_+\\
v^1_-&v^2_-&v^3_-&v^4_-\\
v^2_-&v^1_-&v^4_-&v^3_-\\
\end{array}\right),
\end{equation}
{we can finally write the compact expression}
\begin{equation}
\label{eq:polaritons}
\begin{array}{lll}
\hat{\vec{p}}&=&P\cdot\hat{\vec{O}},
\end{array}
\end{equation}
which greatly simplifies the calculation of the emission rates in terms of the $P_{ij}$ matrix elements.
We recall that, as in the previous sections, with regard to the Jaynes-Cummings polaritons, also the polariton modes of \Eq{eq:fullp} do depend on $N$, a feature that will appear in the next section in the calculation of matrix elements between states with different particle number, and it will be exploited to give intuitive estimates on the order of each of such effects.
\subsection{Transition rates in terms of the polariton eigenstates of the full bosonic model}
\label{sec:rates}
While in the main text the rates were derived in perturbation theory for clarity of exposition, it is also straightforward to compute them after exact diagonalization.
Essential for this calculation are the  commutation relations in \Eq{eq:comm} and the definition of the polaritonic excitations in \Eq{eq:fullp} and their matrix form in \Eq{eq:polaritons}.\\
We start calculating the single-polariton states matrix elements, $M_{G\pm}^n$. We have
\begin{equation}
 {\renewcommand{\arraystretch}{2}
 \begin{array}{lll}
M_{G+}^n&=&\bra{+_{N-1}}(c_{1,n}+c_{2,n})\ket{G_N}
%\\&=&
=\bra{G_{N-1}}\tilde{p}_\pm^{N-1} (c_{1,n}+c_{2,n})\ket{G_N}
%\\&\simeq&
\simeq-\displaystyle{P_{23}}\bra{G_{N-1}} [b^\dagger,c_{2,n}]\ket{G_N}
%\\&=&
=-\displaystyle\frac{P_{23}}{\sqrt{N}},\\
M_{G-}^n&=&\bra{-_{N-1}}(c_{1,n}+c_{2,n})\ket{G_N}\simeq-\displaystyle\frac{P_{43}}{\sqrt{N}},

\end{array}}
\label{mppolar}
 \end{equation}
where the symbol $\simeq$ implies keeping only the lowest non-trivial order in the asymptotic expansion in $1/N$, and where we made explicit the dependence of the polariton mode operator on the number of single-excited fermionic sites, $N$, writing
$\tilde{p}_\pm=\tilde{p}_\pm^{N-1}$ for the mode operator coming from the definition of the bra.
The coefficients $M_{G\pm\pm}^n$ can be computed similarly. However, the lowest non-trivial order in an asymptotic expansion in $1/N$ requires a little bit more work as
\begin{eqnarray}
\sqrt{2}M_{G++}^n&=&\sqrt{2}\bra{++_{N-1}}(c_{1,n}+c_{2,n})\ket{G_N}=\bra{G^{N-1}}{\tilde{p}}^{N-1}_+{\tilde{p}}^{N-1}_+(c_{1,n}+c_{2,n})\ket{G_N}\nonumber\\
&=&\nonumber\displaystyle{\bra{G_{N-1}} ({\tilde{p}}^{N}_+ - \partial_N \tilde{p}^N_+)({\tilde{p}}^{N}_+ - \partial_N \tilde{p}^N_+)(c_{1,n}+c_{2,n})\ket{G_N}}\nonumber\\
&\simeq&\bra{G_{N-1}}\tilde{p}^N_+ \tilde{p}^N_+(c_{1,n}+c_{2,n})\ket{G_N}-\bra{G_{N-1}}\tilde{p}^N_+\partial_N \tilde{p}^N_+(c_{1,n}+c_{2,n})\ket{G_N}%\nonumber\\&&
-\bra{G_{N-1}}\partial_N \tilde{p}^N_+\tilde{p}^N_+(c_{1,n}+c_{2,n})\ket{G_N}\nonumber\\
&\simeq&\nonumber -\frac{1}{\sqrt{N}}\bra{G_{N-1}}\tilde{p}^N_+ (P_{24}c_{2,n}+P_{23}c_{1,n})\ket{G_N}-\displaystyle{\bra{G_{N-1}}{\tilde{p}}^{N}_+(\sum_j\partial_N P_{1j}) P_{jk}^{-1} \hat{p}_k(c_{1,n}+c_{2,n})\ket{G_{N-1}}}\\
&&-\bra{G_{N-1}}\partial_N \tilde{p}^N_+\tilde{p}^N_+(c_{1,n}+c_{2,n})\ket{G_N}\simeq \frac{1}{N}P_{24}P_{23}-\sum_j \partial_N P_{2j}P_{j1}^{-1}
\end{eqnarray}
where, in the second line of the equation above, we transformed 
\begin{eqnarray}
\tilde{p}^{N-1}_+&=&({\tilde{p}}^{N}_+ - \partial_N \tilde{p}^N_+), 
\end{eqnarray}
and, in the third and fourth line of the equation above, we took advantage of the matrix notation of \Eq{eq:polaritons} to evaluate the operator derivative only on the matrix of coefficients $P$, approximating to leading order in $1/N$ the dependence on $N$ only in $g_N$.

Finally, in order to compute the term 
\begin{eqnarray}
\bra{G_{N-1}}{\tilde{p}}^{N}_+(\sum_j\partial_N P_{1j}) P_{jk}^{-1} \hat{p}_k(c_{1,n}+c_{2,n})\ket{G_{N-1}}
\end{eqnarray}
 in the last step of the equation above, we considered that, since we want to calculate the transition matrix elements up to order $O(1/N)$, we could at that point commute the polaritonic operators with the fermionic ones and annihilate the ground state on the right. Hence, the only non-zero contribution in the sum is given by $j=1$ leading to that result. We also used 
 \begin{eqnarray}
 \left(c_{1,n}+c_{2,n}\right)\ket{G_N}&=&\ket{G_{N-1}}=O\left(\frac{1}{\sqrt{N}}\right)
\end{eqnarray} 
  and the fact that since the only dependence on $N$ is through $g_N$, the derivative in $N$ can be written as 
 \begin{equation}
 \partial_N={g_N}/{(2N)}~\partial_{g_N}.
  \end{equation}
Following the same steps, we can calculate the rates towards the other two-polariton states. We obtain
\begin{equation}
\label{eq:Bos_primeprime_2}\ \begin{array}{lll}
M_{G++}^n&=&\displaystyle\frac{P_{24}P_{23}}{\sqrt{2}N}-\displaystyle \frac{g_N{\sum_{j}\partial_{g_N}(P_{2j})P^{-1}_{j1}}}{2\sqrt{2}N},\\
M_{G--}^n&=&\displaystyle\frac{P_{44}P_{43}}{\sqrt{2}N}-\displaystyle \frac{g_N{\sum_{j}\partial_{g_N}(P_{4j})P^{-1}_{j3}}}{2\sqrt{2}N},\\
M_{G+-}^n&=&\displaystyle\frac{P_{24}P_{43}}{N}-\displaystyle \frac{g_N{\sum_{j}\partial_{g_N}(P_{2j})P^{-1}_{j3}}}{2N},\\
&=&\displaystyle\frac{P_{44}P_{23}}{N}-\displaystyle \frac{g_N{\sum_{j}\partial_{g_N}(P_{4j})P^{-1}_{j1}}}{2N},
\end{array}
\end{equation}
which in this form shows clearly that, even in the full-bosonic model, the double-polariton transition rates have a ${1}/{N}$ attenuation with respect to the single-polariton channels, \Eq{mppolar}.
Moreover, the form of \Eq{eq:Bos_primeprime_2} clearly hints at the interpretation of \emph{the electron-subtraction processes as an effective non-adiabatic modulation of the light-matter coupling}, hence allowing to draw a connection with previous experiments in which such process was investigated \cite{Gunter09,Scalari12,Scalari13}. 
 %%%%%
 %%%%%

 %%%%%
 %%%%%
%

\subsection{Photon-scattering transition matrix elements}
\label{phh}
In a real quantum device, the photon emission rate cannot have a perfect efficiency conversion from cavity polariton to emitted light. This quantity is the one measured in photo-detection in a spectroscopic experiment. The photonic emission rate arising from GSE is the product of two processes: First, there is the polariton scattering due to the extraction of an electron, calculated using $ \Gamma_\text{GSE}$, with
 \begin{eqnarray}
 \Gamma_\text{GSE}&=&{\sum_{E=\{\pm,\pm\pm,\pm\mp,\dots\}}}\Gamma^{G\rightarrow E}_\text{el},
\label{eq:gseem}
 \end{eqnarray}
and which we have shown to be dependent on the $\ket{G_N}\rightarrow|\pm_{N-1}\rangle$ channel. 
Then there is a second relaxation process that involves the emission of a photon, 
$|\pm_{N-1}\rangle\rightarrow|G_{N-1}\rangle$, occurring with a probability $|\alpha_\mathrm{ph}^{\pm}|^2$, proportional to the Hopfield coefficients associated with light. 

Note that in $ \Gamma_\text{GSE}$ and elsewhere in rates, in the subscripts of matrix elements and superscripts of rate emissions we omit defining the number of electrons in the initial and final states, where the transition is always $\ket{G_N}\rightarrow\ket{E_{N-1}}$, and in the sum we omit the transition $\ket{G_N}\rightarrow\ket{G_{N-1}}$, which leads only to a \emph{dark electron current} with no photon emission.

Here we calculate explicitly $|\alpha_\mathrm{ph}^{\pm}|^2$, the probability of photon emission associated to the decay of the system from the polariton branches, $\ket{\pm_{N-1}}$, to the ground state ($\ket{G_{N-1}}$ for the RWA case, $\ket{\tilde{G}_{N-1}}$ in the non-RWA bosonic case). This is a fast process occurring \emph{after} the electron scattering process quantified in the sections above. We can calculate $|\alpha_\mathrm{ph}^{\pm}|^2$ in the RWA boson model from 
\begin{eqnarray}
|\alpha_\mathrm{ph}^{\pm}|^2&=&|\bra{+_{N-1}}(a+a^\dagger)\ket{G_{N-1}}|^2
%\\&=&
=|(\alpha^\pm_a)^*+\alpha^\pm_a\frac{\beta_{\pm\pm}}{\sqrt{2}}|^2.
\label{eq:Mgpm2axx}
\end{eqnarray}

%%%%%%%%
In the polariton ground state of the non-RWA Hamiltonian we also have a photonic component for that state, which complicates the calculation. 
Yet, for any fixed $N$, it is always valid the quantum harmonic oscillator relation $\tilde{p}_{\pm}\ket{\tilde{G}}=0$, so all we need to compute is the commutator
\begin{eqnarray}
\lbrack\tilde{p}_{\pm,N},(a+a^\dagger)
\rbrack
&=&v_2^\pm-v_1^\pm
\label{eq:Mgpmfull3}
\end{eqnarray}
and obtain, for the full boson case (non-RWA)
\begin{eqnarray}
|\alpha_\mathrm{ph}^{\pm}|^2=
|\bra{\tilde{G}_{N-1}}\tilde{p}_{\pm,N-1}(a+a^\dagger)\ket{\tilde{G}_{N-1}}|^2&=&|v_1^\pm-v_2^\pm|^2.%\nonumber\\
\label{eq:Mgpmfull4}
\end{eqnarray}
As mentioned in the main text, this calculation allows us to derive the effective total photon emission rate, $\Gamma_\mathrm{tot}=\Gamma^{+}_\mathrm{tot}+\Gamma^{-}_\mathrm{tot}$.

\begin{figure*}[htb!]
\includegraphics[width=16cm]{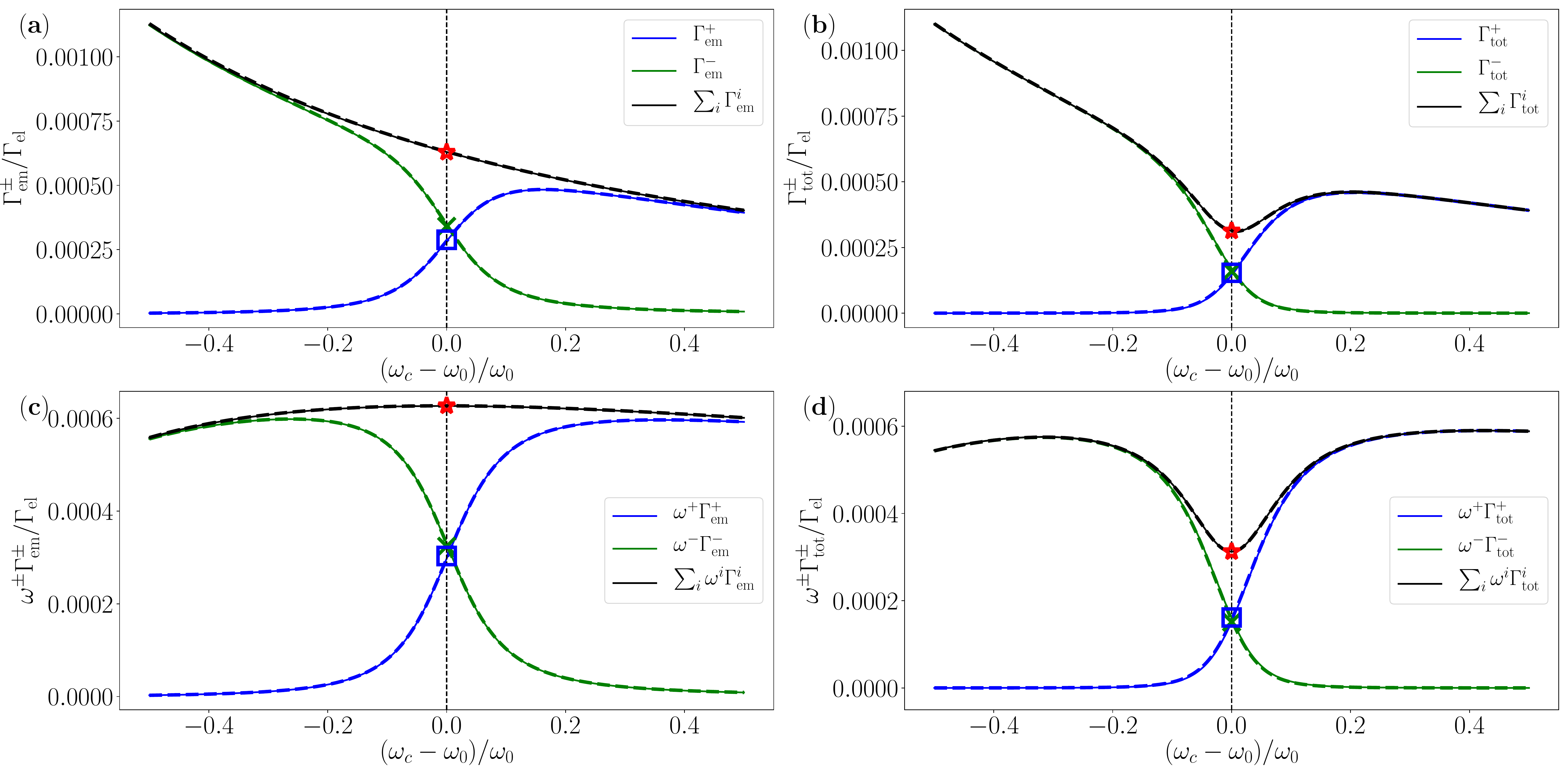}
\centering
\caption{(a,c) Polariton scattering rates $\Gamma^{B'}_{\text{em}}$ and fluxes, $\omega^{B'}\Gamma^{B'}_{\text{em}}$, in units of the total electron transport rate $\Gamma_\text{el}$ for the upper polariton ($B'={+}$, blue curves) and lower polariton ($B'={-}$, green curves), and sum of the two signals (black curves), as a function of the normalized detuning for $g/\omega_0=0.05$. (b,d) Total photon emission rates and fluxes. 
The resonance condition is marked by a dashed black vertical line and markers for the different quantities: blue open square for the upper polariton, green cross for the lower polariton, red star for the total signal. Solid curves correspond to the bosonic RWA quantities, dashed curves to the full boson model developed in the SM.
\label{fig2bis}}
\end{figure*}

\begin{figure*}[htb!]
\includegraphics[width=16cm]{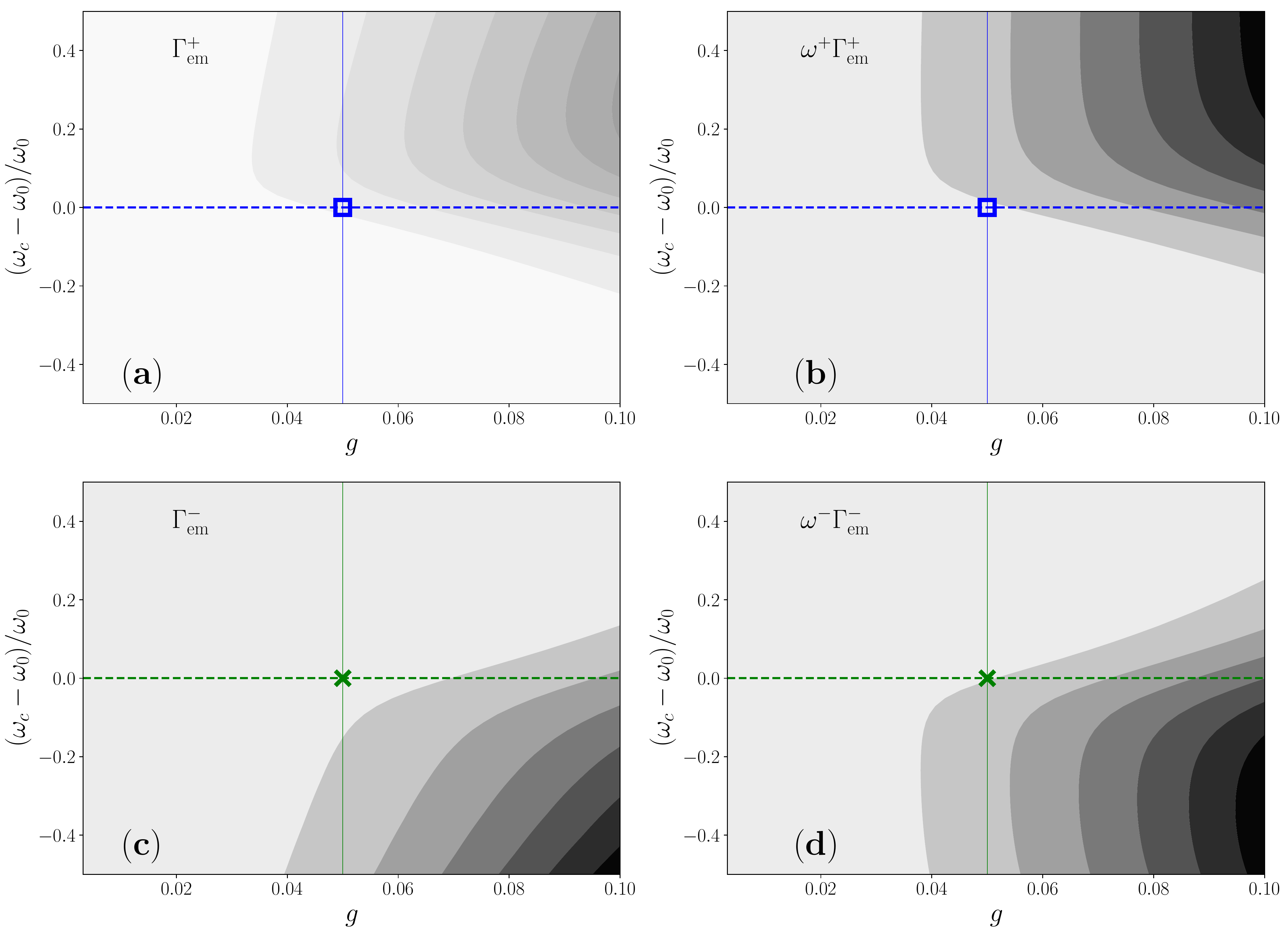}
\centering
\caption{Polariton emission rate $\Gamma^{B'}_{\text{em}}$ [panels (a,c)] and flux [panels (b,d)] as a function of the 
frequency detuning and coupling strength, setting $\chi=3\cdot10^{-3}\omega_0$ fixed and varying $N$, on a logarithmic scale, and thus varying $g=\sqrt{N}\chi$, up to $g=0.1\omega_0$. 
The vertical solid blue and green lines correspond to cut shown Figure~\ref{fig2} in the main text. 
The blue  open square and green cross correspond to the resonance point at $g=0.05\omega_0$ for the upper and lower polariton, respectively. 
The contour plots of panels (a,c) are normalized by a common quantity, as well as those of panels (b,d), showing that the lower polariton rate reaches higher values at fixed $g$, limited in the case of fluxes, $\omega\Gamma_{\text{em}}$. 
\label{fig34s}}
\end{figure*}

The efficiency of this conversion is determined by the cavity characteristic rate, $\Gamma_\text{cav}$, and by the rate of conversion of the bright polaritons into dark polaritons, $\Gamma_\text{dark}^\pm$. If we assume $\Gamma_\text{cav}\gg\Gamma_\text{dark}^\pm$, we obtain  
\begin{eqnarray}
\Gamma^{\pm}_\mathrm{tot}&=&|\alpha_\mathrm{ph}^{\pm}|^2\frac{\Gamma_\mathrm{em}^\pm\Gamma_\mathrm{cav}}{\Gamma_\text{dark}^\mathrm{\pm}+\Gamma_\mathrm{cav}}\simeq|\alpha_\mathrm{ph}^{\pm}|^2\Gamma_\mathrm{em}^\pm.
\end{eqnarray}
Note that $|\alpha_\mathrm{ph}^{\pm}|^2$ is included in the plots in Figure~2 of the main text, here reproduced for clarity as Figure~\ref{fig2bis}, to estimate the effective light emission of the GSE process in both the RWA and non-RWA bosonic models. In Figure~\ref{fig34s}, we resolve the total signals of Figure~\ref{fig33}, into the contributions of the upper and lower polariton branches, shown in the upper and lower panels, respectively. 
For different coupling strengths, the lower (upper) polariton emission rates and fluxes are peaked at negative (positive) frequency detuning. As shown by the darker shading in the contour plots of Figure~\ref{fig34s}, the lower polariton signal is stronger than the upper polariton one, especially at stronger light-matter couplings. 

\begin{figure*}[htb!]
\includegraphics[width=16cm]{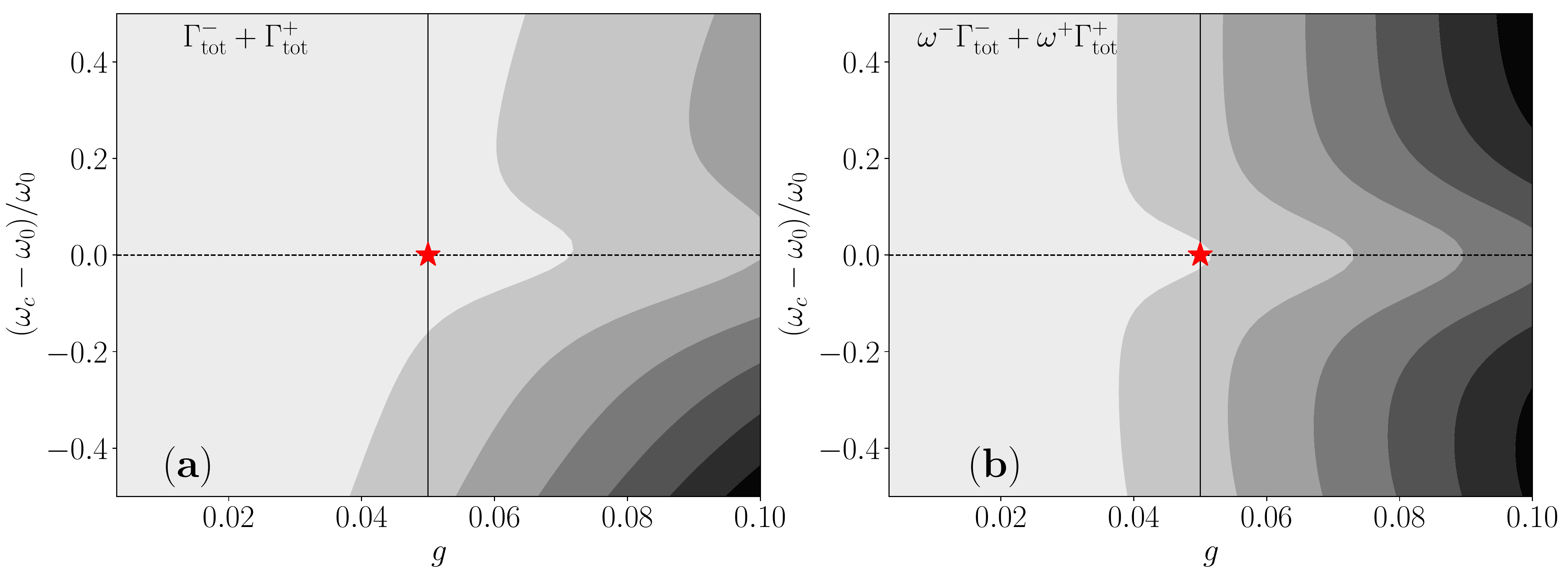}
\centering
\caption{Total extra-cavity photon emission rate $\Gamma_{\text{tot}}=\Gamma^{-}_{\text{tot}}+\Gamma^{+}_{\text{tot}}$ [panel (a)] and flux, $\sum_{B'}\omega^{B'}\Gamma^{B'}_{\text{tot}}$[panel (b)], in units of the total electron scattering rate $\Gamma_\text{el}$. 
\label{fig4s}}
\end{figure*}

\begin{figure*}[htb!]
\includegraphics[width=16cm]{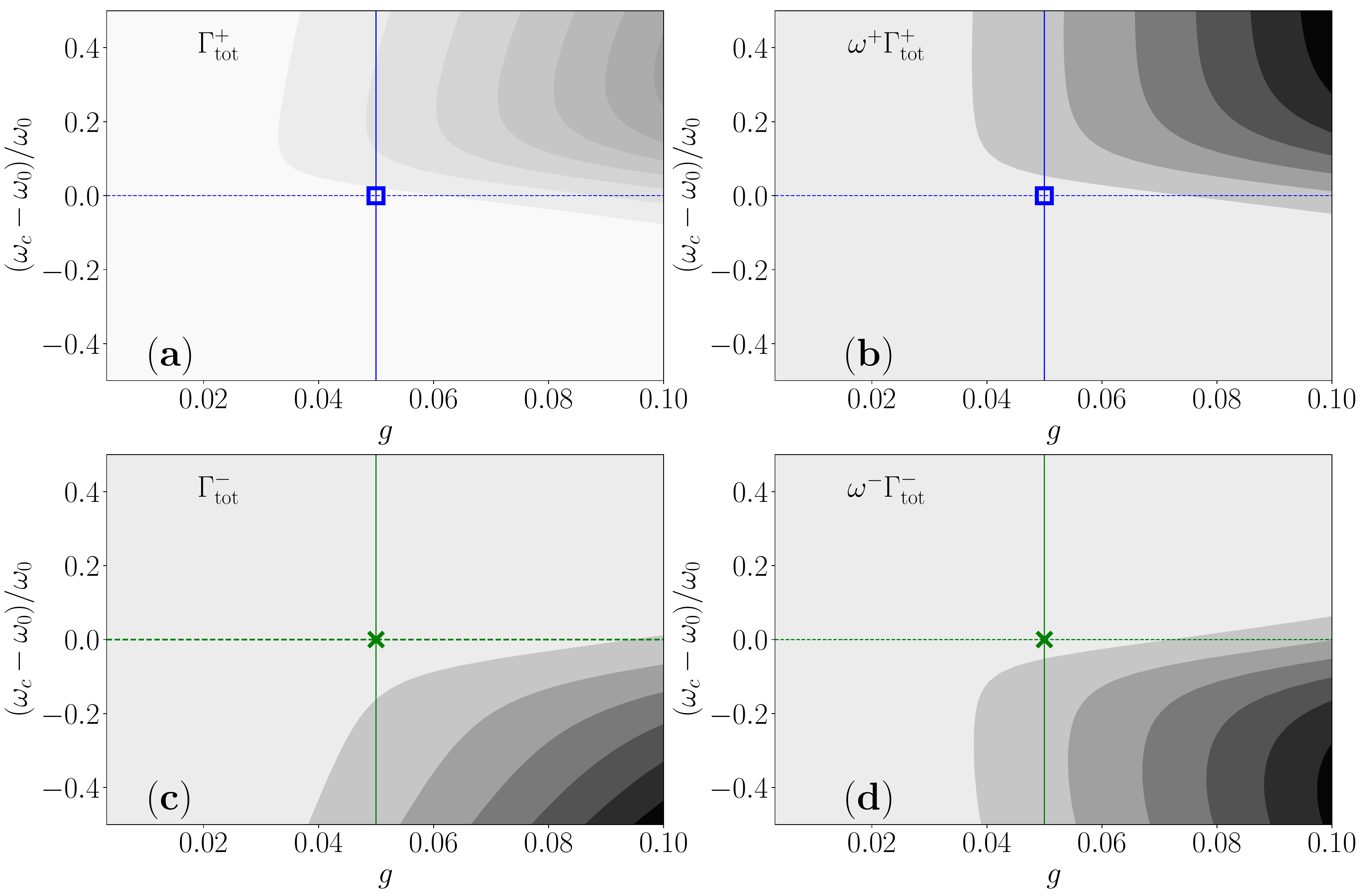}
\centering
\caption{Total extra-cavity photon emission rate $\Gamma^{B'}_{\text{tot}}$ [panels (a,c)] and flux [panels (b,d)], for the upper and lower polariton, as a function of the 
frequency detuning and coupling strength, setting setting $\chi=3\cdot10^{-3}\omega_0$ fixed and varying $N$ and thus $g=\sqrt{N}\chi$. 
The vertical solid blue and green lines correspond to the parameter ranges of the plots of Figure~2 in the main text. 
The blue open square and green cross correspond to the resonance point at $g=0.05\omega_0$ for the upper and lower polariton, respectively.  
\label{fig44b}}
\end{figure*}

In Figure~\ref{fig4s} and Figure~\ref{fig44b}, we report the contour plots for the total extra-cavity emission, showing that the extraction of photons from the GSE is efficient. In particular in Figure~\ref{fig4s} it is visible that this second decay process makes the intensity of the total extra-cavity flux symmetric with respect to detuning of the cavity-matter frequency.    

\section{Fermionic model: Perturbative eigenstates}
\label{sec:appD}
We now consider the full model in \Eq{eq:simplified}. By neglecting the counter-rotating terms which do not conserve the number of bare excitations, we can write the Dicke model in the rotating wave approximation (RWA), i.e. the Tavis-Cummings (TC) model,
\begin{equation}
\label{eq:HRWA}
\begin{array}{lll}
H^\text{RWA}&=&\omega_{c}a^\dagger a+\omega_0 S^3 +\chi(aS^{+}+a^\dagger S^-)+E_0(N,N_2).
\end{array}
\end{equation}
The eigenstates of \Eq{eq:HRWA} can be written in terms of the bare-basis states
\begin{equation}
\label{eq:psi}
\ket{\psi}=\ket{j,m}\otimes\ket{N_2}\otimes\ket{\gamma},
\end{equation}
where $\ket{j,m}=\ket{j,m}_{N}$ is the Dicke state of $N$ two-level systems with total angular momentum $j$ and third component $m$, defined onto the partition of electronic sites containing single excitations, $\ket{N_2}$ counts the double-occupied electron states, which cannot be accounted for in the spin basis, and $\ket{\gamma}$ counts $\gamma$ photons in the photonic subspace.

While \Eq{eq:psi} is in an intuitive form and can be used to find the states diagonalizing \Eq{eq:HRWA}, it will be more convenient to use a different notation, which will be more transparent when the electron-scattering terms of the interaction Hamiltonian will be considered, explicitly showing the dependence on $N$ and $N_2$,
\begin{equation}
\ket{\psi}=\ket{j,m;N,N_2,\gamma},
\label{eq:psi2}
\end{equation}
where the dependence on $\psi$ of the quantum numbers $j$, $m$, $N$, $N_2$, and $\gamma$ will be be highlighted when necessary.
We thus express the eigenstates of the TC Hamiltonian, \Eq{eq:HRWA}, as
\begin{equation}
\ket{\beta^{(0)}}=\sum_{m\gamma}U^{(0)\beta}_{m\gamma}\ket{j^\beta,m;N^\beta,N^\beta_2,\gamma},
\end{equation}
where the factor $U^{(0)\beta}_{m\gamma}$ can be written as
\begin{eqnarray}
U^{(0)\beta}_{m\gamma}&=&u^{(0)\beta}_\gamma\delta_{m, m^\beta_\gamma},
\label{u0def}
\end{eqnarray}
in terms of a $\left(n^\beta+1\right)$-dimensional eigenvector $u^\beta$ of the RWA Hamiltonian in the subspace with $n^\beta=j^\beta+m+\gamma$ bare excitations (derived explicitly in Sec.~\ref{app:RWA}) and where we defined $m^\beta_\gamma=-j^\beta+n^\beta-\gamma$.
We will further denote the energies of the associated states by $E^{(0)}_\beta$.

Using perturbation theory in the potential $V=\chi(S^+ a^+ + S^- a)$, it is now possible to compute the coefficients $U^{\beta}_{m\gamma}$ for the full Hamiltonian of the Dicke model. To first order,
\begin{equation}
\begin{array}{lll}
\ket{\beta}&=&\ket{\beta^{(0)}}-\displaystyle\sum_{\bar{\beta}}\frac{\bra{\bar{\beta}^{(0)}}V\ket{\beta^{(0)}}}{\Delta E_{\beta\bar{\beta}}}\textstyle\ket{\bar{\beta}^{(0)}}
%\\&=&
=\displaystyle\sum_{m\gamma}U^{\beta}_{m\gamma}\ket{j^\beta,m;N^\beta,N^\beta_2,\gamma},
\end{array}
\end{equation}
where $\Delta E_{\beta\bar{\beta}}=E^{(0)}_{\bar{\beta}}-E^{(0)}_\beta$, and we introduced the matrix elements $U^{\beta}_{m\gamma}$.
Equivalently, we can write
\begin{equation}
\begin{array}{lll}
U_{m\gamma}^\beta&=&u^{(0)\beta}_\gamma \delta_{m,m^\beta_\gamma}-\displaystyle\sum_{\bar{\beta}}(c_+^{\beta\bar{\beta}}+c_-^{\beta\bar{\beta}})u^{(0)\bar{\beta}}_\gamma\delta_{m,m^{\bar{\beta}}_\gamma},
\end{array}
\end{equation}
where
\begin{eqnarray}
c_\pm^{\beta\bar{\beta}}&=&\frac{\chi}{\Delta E_{\beta\bar{\beta}}}\sum_{\bar{\gamma}}\bar{u}_{\bar{\gamma}\pm1}^{(0)\bar{\beta}}u^{(0)\beta}_{\bar{\gamma}}A^\beta_{\pm,\bar{\gamma}}~
%\nonumber\\&&\times
\delta_{n^{\bar{\beta}},n^\beta\pm 2}~\delta_{j^{\bar{\beta}}j^\beta}\delta_{N^{\bar{\beta}}N^\beta}\delta_{N_2^{\bar{\beta}}N^\beta_2},
\end{eqnarray}
and
\begin{eqnarray}
A^\beta_{+,\gamma}&=&\sqrt{(\gamma+1)(2j^\beta-n^\beta+\gamma)(n^\beta-\gamma+1)},\;\; \\
A^\beta_{-,\gamma}&=&\sqrt{\gamma(n^\beta-\gamma)(2j^\beta-n^\beta+\gamma+1)}.
\end{eqnarray}
The result of this section is thus the formal definition of the eigenstates of the light-matter system, retaining the full fermionic nonlinearity of \Eq{eq:full}, recasting the problem in terms of collective pseudo-spin states.
The calculation of the eigenstates has been reduced to finding the matrix elements $u^{(0)\beta}_\gamma$, a task that is performed in the next subsection.
\subsection{Single-excitation states of the Tavis-Cummings Hamiltonian}
\label{app:RWA}
In this subsection, we derive an explicit expression for the eigenstates of the rotating wave Hamiltonian in Eq.~(\ref{eq:HRWA}).
To do this, we consider sets of basis states with fixed total angular momentum $j$ and relative number of bare excitations $n=j+\langle \left(a^\dagger a + S^3\right)\rangle_\beta=j+\gamma+m$. In this subspace the matrix elements for the Hamiltonian kernel, $\bar{H}^{j,N,N_2,n}$, read
\begin{eqnarray}
\bar{H}^{j,N,N_2,n}_{ik}&=&\bra{j,m_i;N,N_2,\gamma_i}H^\text{RWA}\ket{j,m_k;N,N_2,\gamma_k}\\
&=&\bra{j,m_i;N,N_2,\gamma_i}\left[\omega_{c}a^\dagger a+\omega_0 S^3 +\chi(aS^{+}+a^\dagger S^-)+E_0(N,N_2)\right]\ket{j,m_k;N,N_2,\gamma_k}\\
&=&\left[ \:\omega_{c}\gamma_k+ \omega_0 m_k+E_0(N,N_2)\right]\delta_{\gamma_i,\gamma_k}
+\bra{j,m_i;N,N_2,\gamma_i} \chi(aS^{+}+a^\dagger S^-)\ket{j,m_k;N,N_2,\gamma_k}\\
&=&\left[ \:\omega_{c}\gamma_k+ \omega_0 m_k+E_0(N,N_2)\right]\delta_{\gamma_i,\gamma_k}\nonumber\\&&
+\chi\sqrt{\gamma_k+1}\sqrt{(j+m_k)(j-m_k+1)}\delta_{\gamma_i,\gamma_k+1}
+\chi\sqrt{\gamma_k}\sqrt{(j-m_k)(j+m_k+1)}\delta_{\gamma_i,\gamma_k-1},
\label{mathjnn2n}
\end{eqnarray}
with $\gamma_k=n-k$, $m_k=-j+n-\gamma_k=-j+k$ where $k=0,\dots,n$.
We define $m_k$ dependent on $\gamma_k$ because the total number of total excitations in each subspace is fixed and equal to $n$.
In the previous equation, the Kr\"{o}necker deltas on $m$ quantum number translated into Kr\"{o}necker deltas on $\gamma$.

Using the definitions of $\gamma_j$ and $m_j$, we then immediately find the matrix elements
\begin{equation}
\label{eq:kernel}
\renewcommand{\arraystretch}{1.8}\begin{array}{lll}
\bar{H}^{j,N,N_2,n}_{k-1,k}&=&\chi\sqrt{n-k+1}\sqrt{k(2j-k+1)},\\
\bar{H}^{j,N,N_2,n}_{k,k}&=&(n-k)\omega_c+k\omega_0+\tilde{E}_0(N,N_2),\\
\bar{H}^{j,N,N_2,n}_{k+1,k}&=&\chi\sqrt{n-k}\sqrt{(k+1)(2j-k)},
\end{array}
\end{equation}
 where 
\begin{equation}
 \tilde{E}_0=E_0(N,N_2)-j\omega_0.
\end{equation} 
Since we are interested in perturbations over the eigenstates of the rotating wave Hamiltonian in Eq.~(\ref{eq:HRWA}), we can specialize the general notation allowing the label $\beta$ to specify not only $j^\beta,N^\beta,N_2^\beta$, but also the total number of bare excitations $n^\beta$.

All of the $\left(n^\beta+1\right)$ eigenvectors $u^{(0)\beta}$ of the matrix $\bar{H}^{j^\beta_{~,} N^\beta_{~,} N_{2,}^\beta, n^\beta}$, defined by \Eq{mathjnn2n}, can then be used in \Eq{u0def} to compute the matrix elements $U^{(0)\beta}_{m\gamma}$ in the rotating wave approximation.

We now explicitly compute the eigenstates for the first excited states in the rotating wave approximation.
The Hamiltonian kernel for the excited states with $n=1$ total excitation takes the form
\begin{equation}
\bar{H}^{j,N,N_2,1}=
\left(\begin{array}{cc}
\omega_c+\tilde{E}_0&\chi\sqrt{2j}\\
\chi\sqrt{2j}&\omega_0+\tilde{E_0}
\end{array}\right).
\end{equation}

The corresponding first two excited polaritonic states have the form
\begin{eqnarray}
\ket{\beta'^{(0)}_N}
&=&\cos{\theta^N_{\beta'}}\ket{j,-j+1;N,N_2,0}%\nonumber\\&&
+\sin{\theta^N_{\beta'}}\ket{j,-j;N,N_2,1}
\end{eqnarray}
where $\ket{\beta'_N}=\{\ket{+_N},\ket{-_N}\}$, labels the upper and lower polaritonic modes, respectively, with the prime denoting the odd parity of these states.
We have 
\begin{eqnarray}
\tan{\theta_+}=(-\Delta+\sqrt{4g_N^2+\Delta^2})/{2g_N}, 
\label{tanp}
\end{eqnarray}
$\theta_+=\theta_{-}-\pi/2$ with $\Delta=\omega_0-\omega_c$, and as always $g_N=\chi\sqrt{N}$.
This corresponds to 
\begin{eqnarray}
{U_{m\gamma}^{+_N}}^{(0)}&=&\delta_{m,-j+1}\delta_{\gamma,0}\cos\theta_{+}-\delta_{m,-j}\delta_{\gamma,1}\sin{\theta_{+}}\nonumber
\end{eqnarray}
and 
\begin{eqnarray}
{U_{m\gamma}^{-_N}}^{(0)}&=&\delta_{m,-j+1}\delta_{\gamma,0}\sin{\theta_{-}}-\delta_{m,-j}\delta_{\gamma,1}\cos{\theta_{-}}\nonumber
\end{eqnarray} 
in \Eq{u0def}.

Now that the single-polariton eigenstates involved in the transition matrix elements of the Fermi golden rule rates are explicitly defined, in order to perform the calculation in \Eq{eq:ratesT0},
it suffices to calculate the action of the fermionic operators describing the electron scattering processes. We will assess this problem in the next subsection, in which the Clebsch-Gordan formalism will be extended to a fermionic Hilbert space in second quantization.
\subsection{Macroscopic states and fermionic master equation}
\label{sec:appF1}

The light-matter states of \Eq{eq:psi2} retain the full fermionic nonlinearity and, as shown in the precedent subsections, can be used to diagonalize the light-matter Hamiltonian of the system.
In order to describe all the different \emph{microscopic} states which constitute a basis for the Hilbert space of the full fermionic Hamiltonian, \Eq{eq:full} one needs to consider \emph{sets} for the single-occupied and double-occupied electron sites, $\mathbf{N}$ and $\mathbf{N_2}$, of cardinality $N$ and $N_2$. For each of those sets, and for each total angular momentum $j$ and third component $m$, we should also take into account the presence of $d$ inequivalent irreducible representations of su(2). We will use the label $r=1,\dots,d$ to characterize this degeneracy. As a consequence, the states for the full fermionic model can be defined as
%%%%%%%%%%%%%
\begin{equation}
\label{eq:microsocopic_states}
\ket{\alpha}=\ket{j,m,r;\mathbf{N},\mathbf{N_2},\gamma}
\end{equation}
where $\gamma$ is the number of photons.
The microscopic information contained in these states is much more than what we need to compute macroscopic effects due to the current flowing through the system.
For this reason, to describe \emph{electron scattering} processes in full generality, we are interested in defining equivalence classes of these states only characterized by the cardinalities of these sets $N$ and $N_2$. 
Furthermore, in order to be able to derive a closed analytical form for the rates induced by adding electronic reservoirs to the model, we extend these equivalence classes to all possible inequivalent representations of su(2). The \emph{equivalence classes} can be explicitly defined as
\begin{eqnarray}
\ket{A}&=&\ket{j,m;N,N_2,\gamma}%\nonumber\\&=&
=\{\ket{\alpha}: N^\alpha=N,N^\alpha_2=N_2,j^\alpha=j,m^\alpha=m;\gamma^\alpha=\gamma\}.%\nonumber\\
\label{eq:macroscopic_states}
\end{eqnarray}
In order to find the relation between microscopic and macroscopic rates, it is instructive to write the \emph{microscopic} rate equation for electron scattering
\begin{equation}
\dot{\rho}_\alpha=-\sum_\beta \Gamma^{\alpha\rightarrow\beta}_\text{el}\rho_\alpha+\sum_\beta \Gamma^{\beta\rightarrow\alpha}_\text{el}\rho_\beta,
\end{equation}
and the corresponding \emph{macroscopic} version by summing over the equivalence classes previously defined
\begin{eqnarray}
\dot{\rho}_A&=&\displaystyle\sum_{\alpha\in A}\dot{\rho_\alpha}=\displaystyle\sum_{\alpha\in A}\sum_{B,\beta\in B} \left(\Gamma^{\beta\rightarrow\alpha}_\text{el}\rho_\beta-
\Gamma_\text{el}^{\alpha\rightarrow\beta}\rho_\alpha\right).
%\nonumber\\
\end{eqnarray}
Note here that $\rho_\alpha$ ($\rho_A$) is the density matrix on the microscopic Hilbert space whose states are described in \Eq{eq:microsocopic_states} (\Eq{eq:macroscopic_states}).

To proceed further we need to make a critical assumption.
We suppose that any coherence present in the system is averaged out in the macroscopic representation, as every electron scattering process needs not to be traced to the different microscopic processes.
The density matrix of the system is then proportional to the identity within each set of states which only differ by the representation label $r$, i.e., we suppose that $\rho_\alpha=\rho_{\tilde{\alpha}}/d_{{\alpha}}$, where we denoted with $\tilde{\alpha}=\{j,m,N,N_2,\gamma\}$ the remaining degrees of freedom. This allows to write
\begin{equation}
\label{eq:Rate_Equation}
\begin{array}{lll}
\dot{\rho}_A&=&-\displaystyle\sum_{B}{\sum}'\sum_{r_\alpha,r_\beta}\Gamma^{\alpha\rightarrow\beta}_\text{el}\frac{\rho_{\tilde{\alpha}}}{d_{\alpha}}
%\\&&
+\displaystyle\sum_{B}{\sum}'\sum_{r_\alpha,r_\beta}\Gamma^{\beta\rightarrow\alpha}_\text{el}\frac{\rho_{\tilde{\beta}}}{d_{\beta}},
\end{array}
\end{equation}
where we used the short-hand notation
\begin{equation}
{\sum}':=\sum_{N^\alpha=N^A,N^\alpha_2=N_2^A} \sum_{N^\beta=N^B,N^\beta_2=N_2^B}.
\end{equation}
\subsection{General transition rates in the fermionic model}
\label{sec:appF}
In this subsection we finally evaluate the rates for transitions between the states $\alpha$ and $\beta$ of the system.
By inserting \Eq{eq:Gamma_Gamma_n} in \Eq{eq:Rate_Equation} we see that this task requires to calculate objects of the form

\begin{equation}
\displaystyle\sum_{r_\alpha,r_\beta}\Gamma_{\text{el},n}^{\alpha\rightarrow\beta}\propto Q_n,
\end{equation}
where
\begin{eqnarray}
Q_n&=&\displaystyle\sum_{r_\alpha,r_\beta}|\bra{\alpha}\hat{O}_n\ket{\beta}|^2,
\label{eq:Q}
\end{eqnarray}
where $\hat{O}_n=(c_{1,n}+c_{2,n})$ or its hermitian conjugate. To proceed, we consider how to define a basis within the degeneracies labelled by $r$.
To this goal, we first write explicitly the following standard basis transformation for the Dicke states introduced in \Eq{eq:psi}
\begin{eqnarray}
\label{eq:CB}
\ket{j_1,j_2;J,M}&=&\sum_{m_1,m_2}C^{J,M}_{j_1,m_1;j_2,m_2}\ket{j_1,m_1}\otimes\ket{j_2,m_2},%\nonumber\\
\end{eqnarray}
where $m_i=-j_i,-j_i+1,\cdots,j_i-1,j_i$, for $i=1,2$, and where we introduced the Clebsch-Gordan coefficients $C^{J,M}_{j_1,m_1;j_2,m_2}=\left(\bra{j_1,m_1}\otimes\bra{j_2,m_2}\right){J,M}\rangle$, reported also in Table~\ref{tab:CB}.

For clarity, we also write the inverse transformation which reads
\begin{equation}
\label{eq:CB_i}
\ket{j_1,m_1}\otimes\ket{j_2,m_2}=\sum_J\sum_MC^{J,M}_{j_1,m_1;j_2,m_2}\ket{j_1,j_2;J,M},
\end{equation}
where $J=|j_1-j_2|,\cdots,j_1+j_2$ and $M=-J,\cdots, J$.

 In particular, we will only be using $j_1={1}/{2}$ in \Eq{eq:CB} in which case the Clebsch-Gordan coefficients are explicitly given by
\begin{table}[t!]
\begin{equation*}
{\renewcommand{\arraystretch}{1.8}
\begin{array}{|c|c|c|}
\hline
&J=j_2+\frac{1}{2}&J=j_2-\frac{1}{2}\\
\hline
m_1=\frac{1}{2}&\sqrt{\frac{J+M}{2J}}&-\sqrt{\frac{J-M+1}{2J+2}},\\
\hline
m_2=-\frac{1}{2}&\sqrt{\frac{J-M}{2J}}&\sqrt{\frac{J+M+1}{2J+2}}.\\
\hline
\end{array}
}
\end{equation*}
\caption{\label{tab:CB} Values of the Clebsch-Gordan coefficients $C^{J,M}_{j_1,m_1;j_2,m_2}$ for $j_1={1}/{2}$.}
\end{table}
and the condition $C^{J,M}_{j_1,m_1;j_2,m_2}=0$, if $m_1+m_2\neq M$.\\
As mentioned in Sections \ref{sec:appD}, \ref{sec:appF1}, there can exist inequivalent representations of su(2) with the same quantum numbers $j, m$ and $N$. The basis within each of these representations can be explicitly obtained by iteratively applying \Eq{eq:CB_i}. 
However, the Hamiltonian has the same action in all these inequivalent representations. We would then have a problem in defining which basis should be used in the Fermi Golden rule.
The solution of this problem comes from our previous assumption of describing these degeneracies with a density matrix proportional to the identity.
This led us to consider a sum over $r_\alpha$ and $r_\beta$ in \Eq{eq:Rate_Equation}.
Thanks to this assumption, \Eq{eq:Q} is valid for any basis in this subspace.
In particular, this allows us to fix a preferential order by which electrons populate the system in the calculations without any loss of generality.
Plugging \Eq{eq:CB_i} into \Eq{eq:microsocopic_states} allows to fully specify the degeneracy $r_\alpha$ of a state with unchanged number of photons $\gamma^\alpha$,
\begin{equation}
\label{cgstate}
\ket{\alpha}=\ket{j^\alpha_1,j^\alpha_2; J^\alpha,M^\alpha,r_\alpha; \mathbf{N}^\alpha,\mathbf{N}^\alpha_2,\gamma^\alpha}
\end{equation}
as  a function of $J^\alpha$, $j_2^\alpha$ and the label $r_{\alpha,2}$ for the degeneracy of the representation with total angular momentum $j_2^\alpha$ (we do not need consider the same for $j^\alpha_1=1/2$ since it is the fundamental representation for the additional electron).
Otherwise stated, we have that
$r_\alpha=r(J^\alpha,j^\alpha_2,r_{\alpha,2})$.

With this in mind, we can then consider the following identity
\begin{equation}
{\renewcommand{\arraystretch}{1.8}
\begin{tabularx}{\textwidth}{ssssstb}
\multicolumn{5}{l}{$\ket{X}=({\blue c_{1,n}}+{\red c_{2,n}})\ket{j_1,j_2;J,M,r_X;\mathbf{N},\mathbf{N}_2,\gamma}$}&=&\\
 &\multicolumn{2}{l}{$\bullet ~n\in \mathbf{N}$}&&&&\\
 &&&\multicolumn{2}{l}{$\bullet~ j_2 = J + \frac{1}{2}$}&&\\
 &&&&$\ket{X}$&=&$C^{JM}_{\frac{1}{2},-\frac{1}{2};J+\frac{1}{2},M+\frac{1}{2}}({\blue c_{1,n}}+{\red c_{2,n}})
 \ket{{\frac{1}{2},-\frac{1}{2}}}\otimes \ket{j_2,M+\frac{1}{2},r_{X,2};\mathbf{N}-n,\mathbf{N}_2,\gamma}$\\
 &&&&&&$+C^{JM}_{\frac{1}{2},\frac{1}{2};J+\frac{1}{2},M-\frac{1}{2}}({\blue c_{1,n}}+{\red c_{2,n}})
 \ket{{\frac{1}{2},\frac{1}{2}}} \otimes\ket{j_2,M-\frac{1}{2},r_{X,2};\mathbf{N}-n,\mathbf{N}_2,\gamma}$\\
 &&&&&=&${\blue C^{JM}_{\frac{1}{2},-\frac{1}{2};J+\frac{1}{2},M+\frac{1}{2}}
 \ket{j_2,M+\frac{1}{2},r_{X,2};\mathbf{N}-n,\mathbf{N}_2,\gamma}}$\\
 &&&&&&$+{\red C^{JM}_{\frac{1}{2},\frac{1}{2};J+\frac{1}{2},M-\frac{1}{2}}
 \ket{j_2,M-\frac{1}{2},r_{X,2};\mathbf{N}-n,\mathbf{N}_2,\gamma}}$\\

 &&&\multicolumn{2}{l}{$\bullet ~j_2 = J - \frac{1}{2}$}&&\\
&&&&$\ket{X}$&=&$C^{JM}_{\frac{1}{2},-\frac{1}{2};J-\frac{1}{2},M+\frac{1}{2}}({\blue c_{1,n}}+{\red c_{2,n}})
\ket{{\frac{1}{2},-\frac{1}{2}}}\otimes \ket{j_2,M+\frac{1}{2},r_{X,2};\mathbf{N}-n,\mathbf{N}_2,\gamma}$\\
&&&&&&$+C^{JM}_{\frac{1}{2},\frac{1}{2};J-\frac{1}{2},M-\frac{1}{2}}({\blue c_{1,n}}+{\red c_{2,n}})
\ket{{\frac{1}{2},\frac{1}{2}}}\otimes \ket{j_2,M-\frac{1}{2},r_{X,2};\mathbf{N}-n,\mathbf{N}_2,\gamma}$\\
&&&&&=&${\blue C^{JM}_{\frac{1}{2},-\frac{1}{2};J-\frac{1}{2},M+\frac{1}{2}}
\ket{j_2,M+\frac{1}{2},r_{X,2};\mathbf{N}-n,\mathbf{N}_2,\gamma}}$\\
&&&&&&$+{\red C^{JM}_{\frac{1}{2},\frac{1}{2};J-\frac{1}{2},M-\frac{1}{2}}
\ket{j_2,M-\frac{1}{2},r_{X,2};\mathbf{N}-n,\mathbf{N}_2,\gamma}}$\\

 &\multicolumn{2}{l}{$\bullet ~n\in \mathbf{N}_2$}&&&&\\
&&&&$\ket{X}$&=&$\ket{{\blue\frac{1}{2},\frac{1}{2}}}\otimes \ket{J,M,r_{X};\mathbf{N},\mathbf{N}_2-n,\gamma}+\ket{{\red\frac{1}{2},-\frac{1}{2}}}\otimes \ket{J,M,r_{X};\mathbf{N},\mathbf{N}_2-n,\gamma}$\\
&&&&&=&${\blue C^{J-\frac{1}{2},M+\frac{1}{2}}_{\frac{1}{2},\frac{1}{2};J-,M}
\ket{\frac{1}{2},J;J-\frac{1}{2},M+\frac{1}{2},r(J-\frac{1}{2},J,r_X),\mathbf{N}+n,\mathbf{N}_2-n,\gamma}}$\\
&&&&&&$+{\blue C^{J+\frac{1}{2},M+\frac{1}{2}}_{\frac{1}{2},\frac{1}{2};J,M}
\ket{\frac{1}{2},J;J+\frac{1}{2},M+\frac{1}{2},r(J+\frac{1}{2},J,r_X);\mathbf{N}+n,\mathbf{N}_2-n,\gamma}}$\\
&&&&&&$+{\red C^{J-\frac{1}{2},M-\frac{1}{2}}_{\frac{1}{2},-\frac{1}{2};J,M}
\ket{\frac{1}{2},J;J-\frac{1}{2},M-\frac{1}{2},r(J-\frac{1}{2},J,r_X);\mathbf{N}+n,\mathbf{N}_2-n,\gamma}}$\\
&&&&&&$+{\red C^{J+\frac{1}{2},M-\frac{1}{2}}_{\frac{1}{2},-\frac{1}{2};J,M}
\ket{\frac{1}{2},J;J+\frac{1}{2},M-\frac{1}{2},r(J+\frac{1}{2},J,r_X);\mathbf{N}+n,\mathbf{N}_2-n,\gamma}}$\\

 &\multicolumn{2}{l}{$\bullet ~n\notin \mathbf{N}, \mathbf{N}_2$}&&&&\\
 &&&&$\ket{X}$&=&$0$,\\
\end{tabularx}}
\label{fermionsss}
\end{equation}
where we highlighted in blue (red) the terms arising from the scattering of a lower (upper) fermion.
As it can be seen above, in \Eq{fermionsss} there are four possible results of the action of the fermionic destruction operators at position $n$, depending on the state of the system at site $n$.
If the site $n$ is occupied by a single electron, $n\in\mathbf{N}$, then the population of the electrons effectively coupled to light diminishes by one. The ket becomes the state $\ket{j_2,M\pm\frac{1}{2},r_{X,2};\mathbf{N}-n,\mathbf{N}_2,\gamma}$ by the action of $c_{1,n}$ ($c_{2,n}$), with a different Clebsch-Gordan coefficient depending on the value of $j_2$.
If the site $n$ is occupied by two electrons, $n\in\mathbf{N}_2$, then the population of the electrons effectively coupled to light \emph{increases} by one. The ket then becomes of the form of \Eq{cgstate} and the averaging of the processes involved in the macroscopic state scattering eventually leads to a ket of the form of \Eq{eq:microsocopic_states}.
If the site $n$ is not occupied by any electron, no scattering occurs.

We thus can appreciate how \Eq{fermionsss} is fundamental for the use of Dicke state formalism in the open dynamics considered here, bridging a connection with the second-quantization formalism and which can be used also for the study of other processes. Note that Ref.~\cite{Zhang18b} contains related independent work in first quantization, in the context of superradiant lasing.

By inserting these results in Eq.~(\ref{eq:Q}), we find
\begin{eqnarray}
Q_n&=&\displaystyle\sum_{r_\alpha,r_\beta}|\bra{\beta}(c_{1,n}+c_{2,n})\ket{\alpha}|^2\\
&=&\nonumber\displaystyle\sum_{r_\alpha,r_\beta}\left|\displaystyle\left[\delta_{\mathbf{N}^{{\beta}},\mathbf{N}^\alpha-n}\delta_{\mathbf{N}_2^{{\beta}},\mathbf{N}^\alpha_2}\delta_{r_{\beta},r_{\alpha,2}}
\delta_{{J}^{{\beta}},{J^{\alpha}}+\frac{1}{2}}\right.\right.
\left.\left.\left(\delta_{m_\beta,m+\frac{1}{2}}C^{{J^{\alpha}}m_\alpha}_{{J^{\alpha}}+\frac{1}{2},m_\alpha+\frac{1}{2}}
+\delta_{m_\beta,m_\alpha-\frac{1}{2}}D^{{J^{\alpha}}m_\alpha}_{{J^{\alpha}}+\frac{1}{2},m_\alpha-\frac{1}{2}}\right)\right.\right.\\
\nonumber&&+\delta_{\mathbf{N}^{{\beta}},\mathbf{N}^\alpha-n}\delta_{\mathbf{N}_2^{{\beta}},\mathbf{N}^\alpha_2}\delta_{{J}^{{\beta}},{J^{\alpha}}-\frac{1}{2}}\delta_{r_{\beta},r_{\alpha,2}}
\left.\left(\delta_{m_\beta,m_\alpha+\frac{1}{2}}C^{{J^{\alpha}}m_\alpha}_{{J^{\alpha}}-\frac{1}{2},m_\alpha+\frac{1}{2}}+\delta_{m_\beta,m_\alpha-\frac{1}{2}}D^{{J^{\alpha}}m_\alpha}_{{J^{\alpha}}-\frac{1}{2},m_\alpha-\frac{1}{2}}\right)\right.\\
\nonumber&&+\delta_{\mathbf{N}^{{\beta}},\mathbf{N}^\alpha+n}\delta_{\mathbf{N}_2^{{\beta}},\mathbf{N}^\alpha_2-n}\delta_{{J}^{{\beta}},{J^{\alpha}}-\frac{1}{2}}\delta_{r_{\beta},r(J^\alpha-\frac{1}{2},J^\alpha,r_{\alpha})}
\left.\left(\delta_{m_\beta,m_\alpha+\frac{1}{2}}D^{{J^{\alpha}}-\frac{1}{2}m_\alpha+\frac{1}{2}}_{{J^{\beta}},m_\alpha}+\delta_{m_\beta,m_\alpha-\frac{1}{2}}C^{{J^{\alpha}}-\frac{1}{2}m_\alpha-\frac{1}{2}}_{{J^{\alpha}},m_\alpha}\right)\right.\\
&&+\delta_{\mathbf{N}^{{\beta}},\mathbf{N}^\alpha+n}\delta_{\mathbf{N}_2^{{\beta}},\mathbf{N}^\alpha_2-n}\delta_{{J}^{{\beta}},{J^{\alpha}}+\frac{1}{2}}\delta_{r_{\beta},r(J^\alpha+\frac{1}{2},J^\alpha,r_{\alpha})}
\left.\left.\left.\left(\delta_{m_\beta,m_\alpha+\frac{1}{2}}D^{{J^{\alpha}}+\frac{1}{2}m_\alpha+\frac{1}{2}}_{{J^{\alpha}},m_\alpha}+\delta_{m_\beta,m_\alpha-\frac{1}{2}}C^{{J^{\alpha}}+\frac{1}{2}m_\alpha-\frac{1}{2}}_{{J^{\alpha}},m_\alpha}\right)\right.\right]\right|^2,\nonumber\\\label{cd1}
\end{eqnarray}
which, in turn, allows to perform the two sums over the representation indexes to obtain
\begin{eqnarray}
Q_n&=&d_\alpha\nonumber\displaystyle\left|\displaystyle\left[\delta_{N^{{\beta}},N^\alpha-n}\delta_{N_2^{{\beta}},N^\alpha_2}
\delta_{{J}^{{\beta}},{J^{\alpha}}+\frac{1}{2}}\right.\right.
\left.\left.\left(\delta_{m_\beta,m_\alpha+\frac{1}{2}}C^{{J^{\alpha}}m_\alpha}_{{J^{\alpha}}+\frac{1}{2},m_\alpha+\frac{1}{2}}
+\delta_{m_\beta,m_\alpha-\frac{1}{2}}D^{{J^{\alpha}}m_\alpha}_{{J^{\alpha}}+\frac{1}{2},m_\alpha-\frac{1}{2}}\right)\right.\right.\\
\nonumber&&+\delta_{\mathbf{N}^{{\beta}},\mathbf{N}^\alpha-n}\delta_{\mathbf{N}_2^{{\beta}},\mathbf{N}^\alpha_2}\delta_{{J}^{{\beta}},{J^{\alpha}}-\frac{1}{2}}
\left.\left(\delta_{m_\beta,m_\alpha+\frac{1}{2}}C^{{J^{\alpha}}m_\alpha}_{{J^{\alpha}}-\frac{1}{2},m_\alpha+\frac{1}{2}}+\delta_{m_\beta,m_\alpha-\frac{1}{2}}D^{{J^{\alpha}}m_\alpha}_{{J^{\alpha}}-\frac{1}{2},m_\alpha-\frac{1}{2}}\right)\right.\\
\nonumber&&+\delta_{\mathbf{N}^{{\beta}},\mathbf{N}^\alpha+n}\delta_{\mathbf{N}_2^{{\beta}},\mathbf{N}^\alpha_2-n}\delta_{{J}^{{\beta}},{J^{\alpha}}-\frac{1}{2}}
\left.\left(\delta_{m_\beta,m_\alpha+\frac{1}{2}}D^{{J^{\alpha}}-\frac{1}{2}m_\alpha+\frac{1}{2}}_{{J^{\beta}},m_\alpha}+\delta_{m_\beta,m_\alpha-\frac{1}{2}}C^{{J^{\alpha}}-\frac{1}{2}m_\alpha-\frac{1}{2}}_{{J^{\alpha}},m_\alpha}\right)\right.\\
&&+\delta_{\mathbf{N}^{{\beta}},\mathbf{N}^\alpha+n}\delta_{\mathbf{N}_2^{{\beta}},\mathbf{N}^\alpha_2-n}\delta_{{J}^{{\beta}},{J^{\alpha}}+\frac{1}{2}}
\left.\left.\left.\left(\delta_{m_\beta,m_\alpha+\frac{1}{2}}D^{{J^{\alpha}}+\frac{1}{2}m_\alpha+\frac{1}{2}}_{{J^{\alpha}},m_\alpha}+\delta_{m_\beta,m_\alpha-\frac{1}{2}}C^{{J^{\alpha}}+\frac{1}{2}m_\alpha-\frac{1}{2}}_{{J^{\alpha}},m_\alpha}\right)\right.\right]\right|^2.\label{cd2}
\end{eqnarray}
The coefficients $C$ and $D$ appearing in Eqs.~(\ref{cd1}),(\ref{cd2}) are the Clebsch-Gordan coefficients defined as
\begin{equation}
\begin{array}{lll}
C^{J,M}_{j,m}&=&\left(\delta_{J,j+\frac{1}{2}}\sqrt{\frac{J-M}{2J}}+\delta_{J,j-\frac{1}{2}}\sqrt{\frac{J+M+1}{2J+2}}\;\right)\delta_{M,m-\frac{1}{2}},\\
D^{J,M}_{j,m}&=&\left(\delta_{J,j+\frac{1}{2}}\sqrt{\frac{J+M}{2J}}-\delta_{J,j-\frac{1}{2}}\sqrt{\frac{J-M+1}{2J+2}}\;\right)\delta_{M,m+\frac{1}{2}},
\end{array}
\end{equation}
for which the different coefficients connecting different pseudo-spin states are summarized in Table~\ref{tab:CB}.

Having performed the sum over the representation index allows to write \Eq{eq:Rate_Equation} as
\begin{equation}
\begin{array}{lll}
\dot{\rho}_A&=&-\displaystyle\sum_{B}\sum_{N^\alpha=N^A,N^\alpha_2=N_2^A} \sum_n\sum_{N^\beta=N^B,N^\beta_2=N_2^B}\sum_i\lambda^i_{\alpha\beta}Q_n\frac{\rho_{\mathbf{N}^\alpha,\mathbf{N}^\alpha_2,j_\alpha,m_\alpha,\gamma_\alpha}}{d_{\alpha}}\\
&&+\displaystyle\sum_{B}\sum_{N^\alpha=N^A,N^\alpha_2=N_2^A} \sum_n\sum_{N^\beta=N^B,N^\beta_2=N_2^B}\sum_i\lambda^i_{\alpha\beta}Q_n\frac{\rho_{\mathbf{N}^\beta,\mathbf{N}^\beta_2,j_\beta,m_\beta,\gamma_\beta}}{d_{\beta}},
\end{array}
\end{equation}
where $i=\text{in}, \text{out}$, following the definition inspired by \Eq{eq:rates}, 
\begin{eqnarray}
\lambda^\text{out}_n&=&2\pi\lambda^2
({1-\bar{n}(-\Delta_{\alpha\beta})})\nu(-\Delta_{\alpha\beta})
\end{eqnarray} 
and 
\begin{eqnarray}
\lambda^\text{in}_n&=&2\pi\lambda^2({\bar{n}(\Delta_{\alpha\beta})})\nu(\Delta_{\alpha\beta}). 
\end{eqnarray}
We also arranged the sums in the order we will perform them. We now give the arguments to simplify the calculations on the first line of the above formula. Analogous considerations can be applied to the second line.\\
 For a fixed $\mathbf{N}^\alpha$ and $\mathbf{N}_2^\alpha$, and for each $n$ the sum over $\mathbf{N}^\beta$ and $\mathbf{N}_2^\beta$ can be performed using the deltas in the expression for $Q_n$. The net effect of this operation is to limit the range of the sum over $n$ to those electrons which can be added or removed to turn the sets  $\mathbf{N}^\alpha$ into $\mathbf{N}^\beta$ or $\mathbf{N}^\alpha_2$ into $\mathbf{N}_2^\beta$. Since nothing depends explicitly on such a specific $n$, the sum over $n$ will return the number of microscopioc ways $\kappa_{A\rightarrow B}$ to go from a specific set characterized by $N^A$, $N_2^A$ to one charaterized by $N^B$ and $N^B_2$. The last sum over $\{N^\beta\}$ and $\{N_2^\beta\}$ can now be used to sum over the probability density 
\begin{equation}
 \sum_{N^\alpha=N^A,N^\alpha_2=N_2^A} \rho_{N^\alpha,N^\alpha_2,j^\alpha,m^\alpha} 
 \end{equation}
 to give $\rho_A$ to obtain
\begin{equation}
\begin{array}{lll}
\dot{\rho}_A&=&-\displaystyle\sum_{B}\Gamma_{L/R}^{A\rightarrow B}\rho_A+\displaystyle\sum_{B}\Gamma_{L/R}^{B\rightarrow A}\rho_B,
\end{array}
\end{equation}
where
\begin{eqnarray}
\Gamma^{A\rightarrow {B}}_{L/R,\text{in/out}}&=&\nonumber\theta^{\text{AB}}_{L/R,\text{in/out}}\kappa_{A\rightarrow {B}}
\left|\displaystyle\sum_{\bar{m},m,\gamma}\right.%\nonumber\\&&\nonumber
\left[\delta_{N^{{B}},N^A-1}\delta^{\text{in/out}}_{N_2^{{B}},N^A_2}
\delta_{{J}^{{B}},{J^{A}}+\frac{1}{2}}\right.
\left({U}^{{B}}_{\bar{m}\gamma}\right)^*U^{{A}}_{{m}{\gamma}}\nonumber\\&&\times
\left(\delta_{\bar{m},m+\frac{1}{2}}C^{{J^{A}}m}_{{J^{A}}+\frac{1}{2},m+\frac{1}{2}}+\delta_{\bar{m},m-\frac{1}{2}}D^{{J^{A}}m}_{{J^{A}}+\frac{1}{2},m-\frac{1}{2}}\right)\nonumber\\\nonumber
&&+\delta_{N^{{B}},N^A-1}\delta^\text{in/out}_{N_2^{{B}},N^A_2}\delta_{{J}^{{B}},{J^{A}}-\frac{1}{2}}
\left.\left({U}^{{B}}_{\bar{m}\gamma}\right)^*U^{{A}}_{{m}{\gamma}}\left(\delta_{\bar{m},m+\frac{1}{2}}C^{{J^{A}}m}_{{J^{A}}-\frac{1}{2},m+\frac{1}{2}}+\delta_{\bar{m},m-\frac{1}{2}}D^{{J^{A}}m}_{{J^{A}}-\frac{1}{2},m-\frac{1}{2}}\right)\right.\\
&&\nonumber+\delta_{N^{{B}},N^A+1}\delta^\text{in/out}_{N_2^{{B}},N^A_2-1}\delta_{{J}^{{B}},{J^{A}}-\frac{1}{2}}
\left.\left({U}^{{B}}_{\bar{m}\gamma}\right)^*U^{{A}}_{{m}{\gamma}}\left(\delta_{\bar{m},m+\frac{1}{2}}D^{{J^{A}}-\frac{1}{2}m+\frac{1}{2}}_{{J^{B}},m}+\delta_{\bar{m},m-\frac{1}{2}}C^{{J^{A}}-\frac{1}{2}m-\frac{1}{2}}_{{J^{A}},m}\right)\right.\\
&&+\delta_{N^{{B}},N^A+1}\delta^\text{in/out}_{N_2^{{B}},N^A_2-1}\delta_{{J}^{{B}},{J^{A}}+\frac{1}{2}}
\left.\left.\left.\left({U}^{{B}}_{\bar{m}\gamma}\right)^*U^{{ A}}_{{m}{\gamma}}\left(\delta_{\bar{m},m+\frac{1}{2}}D^{{J^{A}}+\frac{1}{2}m+\frac{1}{2}}_{{J^{A}},m}+\delta_{\bar{m},m-\frac{1}{2}}C^{{J^{A}}+\frac{1}{2}m-\frac{1}{2}}_{{J^{A}},m}\right)\right.\right]\right|^2,\nonumber\\
\label{eq:ratesFermions}
\end{eqnarray}
where 
\begin{eqnarray}
\theta^{\text{AB}}_{L/R,\text{out}}&=&\Gamma_\text{el}[1+\theta(\mu_{L/R}+\Delta_{AB})], \\
\theta^{\text{AB}}_{L/R,\text{in}}&=&\Gamma\theta(\mu_{L/R}-\Delta_{AB}), 
\end{eqnarray}
$\delta^\text{out}_{a,b}=\delta_{a,b}$, and $\delta^{\text{in}}_{a,b}=\delta_{a,b+1}$ and where
\begin{equation}
\begin{array}{lll}
\kappa_{(N,N_2)\rightarrow(N+1,N_2)}&=&N_T-N-N_2,\\
\kappa_{(N,N_2)\rightarrow(N-1,N_2)}&=&N,\\
\kappa_{(N,N_2)\rightarrow (N-1,N_2+1)}&=&N,\\
\kappa_{(N,N_2)\rightarrow (N+1,N_2-1)}&=&N_2,
\end{array}
\end{equation}
where $N_T$ is the number of total electron sites.

While Eq.~(\ref{eq:ratesFermions}) contains all the information needed to compute the transition rates, it is worth to proceed a little further to obtain an expression which more clearly highlights its physical content. To achieve this, we decompose the coefficients $U^B_{m\gamma}$ as
\begin{equation}
U^B_{m\gamma}=\sum_n u^B_\gamma(n)\delta_{m,-j^B+n-\gamma},
\end{equation}
where the coefficients $u^B_\gamma(n)$ account for the coherences of the state $B$ within the subspace with $n$ bare excitations.
For example, at first order in the perturbation potential $V$, we have
\begin{equation}
\label{eq:UUU}
\begin{array}{lll}
u^B_{\gamma}(n^B)&=&u^{(0)B}_\gamma\\
u^B_\gamma(n^B\pm 2)&=&-\displaystyle\sum_{\bar{B}}c_{\pm}^{B\bar{B}}u^{(0)\bar{B}}_\gamma,
\end{array}
\end{equation}
while all other coefficients are zero. Note that the range of $\gamma$ is constrained inside each $u^{(0)B}_\gamma$ so that, for each $n$, we have $0\leq\gamma\leq n$. However, both Eq.~(\ref{eq:ratesFermions}) and Eq.~(\ref{eq:UUU}) are general and can, in principle, be applied to higher order cases. Using this notation in Eq.~(\ref{eq:ratesFermions}) leads to quantities of the form
\begin{equation}
\bar{Q}=\displaystyle\sum_{m,\bar{m},\gamma}\left({U}^{B}_{\bar{m},m}\right)^*U^{A}_{m,\gamma}\delta_{\bar{m},m+a}\delta_{j^B_,j^A+b}F(m),
\end{equation}
where $a,b\in\mathbb{R}$ and $F$ a generic function. We have
\begin{equation}
\begin{array}{lll}
\bar{Q}&=&\displaystyle\sum_{n,\bar{n}}\sum_{m,\gamma}\bar{u}^{B}_\gamma(\bar{n})u^{A}_{\gamma}(n)F(m)
%\\&&\times
\delta_{m+a,-j^B+\bar{n}-\gamma}\delta_{m,-j^A+n-\gamma}\delta_{j^B_,j^A+b}\\
&=&\displaystyle\sum_{n,\gamma}\bar{u}^{B}_\gamma(n+a+b)u^{A}_\gamma(n)F(-j^A+n-\gamma)\delta_{j^B_,j^A+b}%\\&=&
=\delta_{j^B_,j^A+b}\langle A,B\rangle_{F}^{a+b},
\end{array}
\end{equation}
where we defined the following pseudo-inner product
\begin{equation}
\langle A, B \rangle^{x}_F=\sum_n\sum_\gamma \bar{u}^{B}_\gamma (n+x) u^{A}_\gamma(n)F(-j^A+n-\gamma),
\end{equation}
which quantifies an effective overlap between coherent components of $A$ and $B$ which belongs to subspaces having total number of bare excitations which differ by $x$. With this notation at hand, we can rewrite Eq.~(\ref{eq:ratesFermions}) as
\begin{equation}
\label{eq:ratesInner_app}
\begin{array}{l}
\Gamma^{A\rightarrow {B}}_{L/R,\text{in/out}}=\theta^{\text{AB}}_{L/R,\text{in/out}}\kappa_{A\rightarrow {B}}
%\\\times
\left|\left[\delta_{\Delta N,-1}\delta^{\text{in/out}}_{N_{2,}^{{B}}N^A_2}\delta_{\Delta^N_J,-1}\left(\langle A,B\rangle^{1}_{C^{\downarrow}_{AB}}+\langle A,B\rangle^{0}_{D^{\downarrow}_{AB}}\right)\right.\right.\\
+\delta_{\Delta N,-1}\delta^{\text{in/out}}_{N_{2,}^{{B}}N^A_2}\delta_{\Delta^N_J,0}\left(\langle A,B\rangle^{0}_{C^{\downarrow}_{AB}}+\langle A,B\rangle^{-1}_{D^{\downarrow}_{AB}}\right)%\\
+\delta_{\Delta N,1}\delta^{\text{in/out}}_{N_{2,}^{{B}}N^A_2-1}\delta_{\Delta^N_J,0}\left(\langle A,B\rangle^{1}_{D^{\uparrow}_{BA}}+\langle A,B\rangle^{0}_{C^{\uparrow}_{BA}}\right)\\

+\left.\left.\delta_{\Delta N,1}\delta^{\text{in/out}}_{N_{2,}^{{B}}N^A_2-1}\delta_{\Delta^N_J,1}\left(\langle A,B\rangle^{0}_{D^{\uparrow}_{BA}}+\langle A,B\rangle^{-1}_{C^{\uparrow}_{BA}}\right)\right]\right|^2,\\
\end{array}
\end{equation}
where $C^\downarrow_{AB}(m)=C^{j^A,m}_{j^B,m+\frac{1}{2}}$, $C^\uparrow_{AB}(m)=C^{j^A,m-\frac{1}{2}}_{j^B,m}$, $D^\downarrow_{AB}(m)=D^{j^A,m}_{j^B,m-\frac{1}{2}}$, $D^\uparrow_{AB}(m)=D^{j^A,m+\frac{1}{2}}_{j^B,m}$. We also defined $\Delta N=N_B-N_A$ and changed the notation to highlight the quantity 
\begin{equation}
\Delta^N_J=(j_{N^B}-j^B)-(j_{N^A}-j^A)=(N^B-N^A)/2-(j^B-j^A),
\end{equation}
 which identifies changes in the symmetry of the state. For example, $\Delta^N_J<0$ indicates a transition towards a state more symmetric than the original one.\\
This notation highlights some of the physical content of these expressions. For example, transitions between states with equal (different) parity take contributions only from factors proportional to $\langle A, B\rangle^n$ with an even (odd) $n$.
\subsection{Emission rate for ground state electroluminescence}
\label{sec:appG}
In this subsection we analytically estimate the population for the lowest energy states as current passes through the system. Each of the results will be given at lowest non-trivial order in the normalized light-matter coupling $\eta=g_{\bar{N}}/\omega_0$ and the light-matter coupling $\eta/\omega_0$. To simplify the analysis we will further assume that $N_2=0$, which allow a closed analytical treatment.
Hereafter, the notation $B'$ ($B''$) will indicate one-polariton (two-polariton) eigenstates.

Essential in the discussion, the expression for the ground state with $N$ electrons is identified by
\begin{equation}
\label{eq:gs}
u^{G_N}_{\gamma}(n)=\delta_{\gamma,0}\delta_{n,0}-\delta_{n,2}\sum_{\bar{B}''} \frac{g_N}{\Delta E_{B''}}\bar{u}^{(0)B''}_1 u^{(0)\bar{B}''}_\gamma,
\end{equation}
This formula shows explicitly that the ground state is a coherent superposition of states with different number of bare excitations. These are the coherences necessary for {ground state} electroluminescence.\\

\begin{figure*}[t]
\includegraphics[width=12cm]{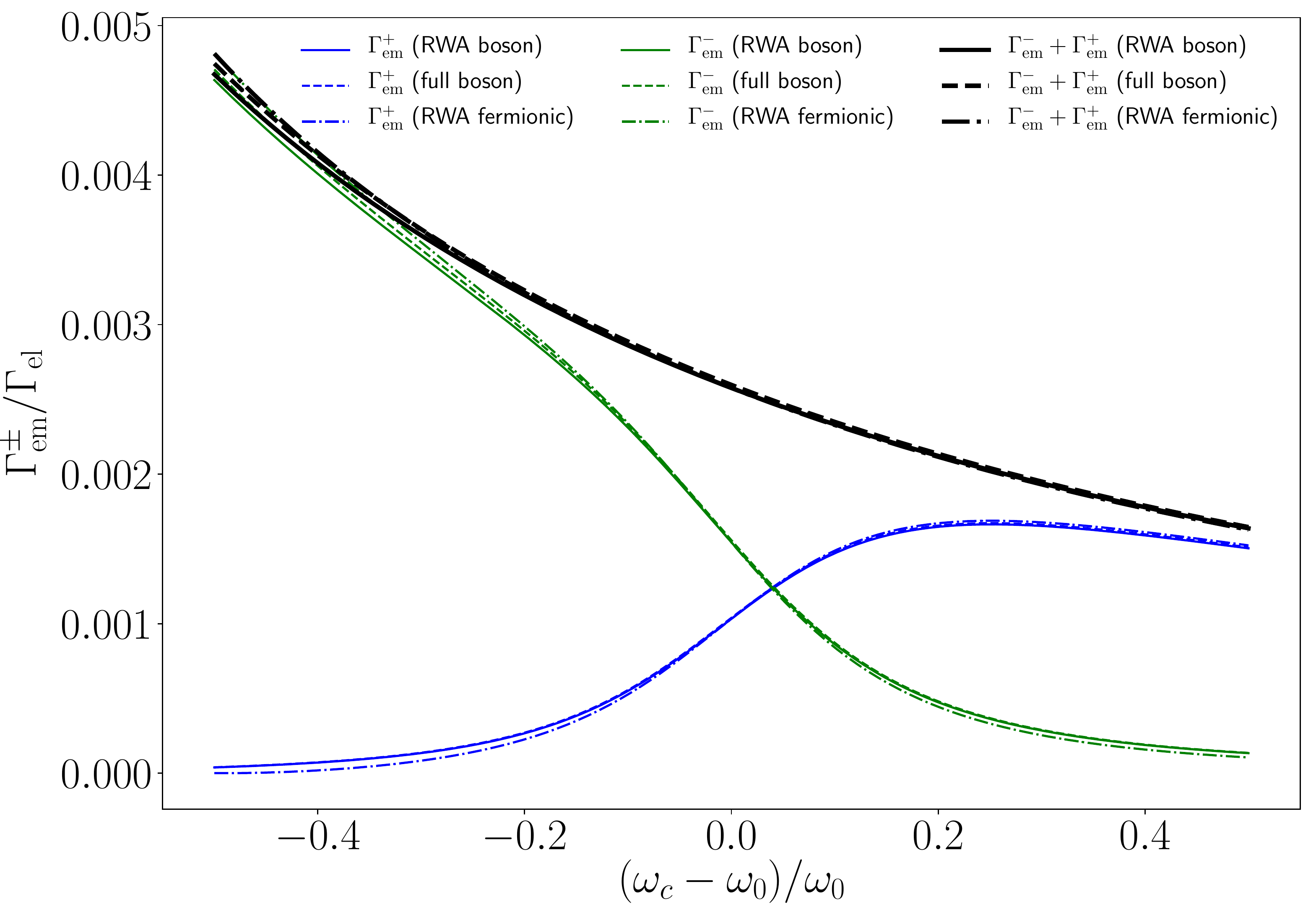}
\centering
\caption{\label{figappfermionsingle}
{\bf Single-polariton emission rates in the fermionic and bosonic models}. Emission rates $\Gamma^{B'}_{\text{em}}$ in units of the emission rate $\Gamma_\mathrm{el}$ for $B'={+}$ (monotonically increasing blue curves) and $B'={-}$ (monotonically decreasing green curves), for the upper and lower polariton, respectively, and the sum of the two contributions (black thick curves) as a function of the normalized detuning $(\omega_c-\omega_0)/\omega_0$ for $g_{\bar{N}}/\omega_0=0.1$, with $\Delta=\omega_0-\omega_c$.
The solid curves correspond to the perturbative bosonic model, in terms of the Jaynes-Cummings polaritons, used in the main text.
The dashed curves correspond to the full bosonic model, where the Hamiltonian is directly diagonalized.
The dot-dashed curves represent the rates obtained from the fermionic model, with a perturbative calculation of the eigenstates in terms of the fermionic RWA Hamiltonian.
The bosonic approximations fit very well the fermionic model to an extent such that the two curves are barely distinguishable in the scale used for the plot, up to large detuning.}
\end{figure*}

The virtual polaritonic population present in the ground state highlighted in Eq.~(\ref{eq:gs}) allows for new non-zero out-rates. As a consequence, in the case $\Gamma_\text{cav}\gg\Gamma_\text{el}$, we will estimate the total emission rate from the decay of the polariton $B=B',B''$ using the shorthand notation
 \begin{equation}
 \begin{array}{lll}
 \langle G,B'\rangle&=&\langle G_N,B'_{N-1}\rangle^{-1}_{D^{\downarrow}_{G_N B'_{N-1}}},\\
 \langle G,B''\rangle&=&\langle G_N,B''_{N-1}\rangle^{0}_{C^{\downarrow}_{G_NB''_{N-1}}},
 \end{array}
 \end{equation}
where as elsewhere, $B'$ and $B''$ indicate the one-polariton and two-polariton eigenstates, respectively.
For clarity, we stop a moment to anticipate some future results. The behaviour of the pseudoinner products in the previous expression are, at lowest non-trivial order, given by  
\begin{eqnarray}
\langle G,B'\rangle\propto D^\downarrow_{G_N,B_{N-1}}=O\left(\frac 1 N \right) 
\end{eqnarray}
and 
\begin{eqnarray}
 \langle G,B''\rangle\propto \partial_N g_N ~C^\downarrow_{G_N,B''_{N-1}} =O\left(\frac {1}{ N^2} \right) .
 \end{eqnarray}
More explicitly, the pseudo-inner products can be calculated as
\begin{eqnarray}
\langle G_N,B_{N-1}'\rangle&=&\displaystyle\sum_{n,\gamma}\bar{u}^{B'_{N-1}}_\gamma(n-1)u^{G_N}_{\gamma}(n) D^{j^{G_N},m^{G_N}_\gamma}_{j^{G_N}-\frac{1}{2},m^{G_N}_{\gamma,n}-\frac{1}{2}}%\nonumber\\&=&
=\displaystyle\sum_{\gamma=0}^1\bar{u}_\gamma^{B'_{N-1}}(1)u^{G_N}_\gamma(2)\sqrt{\frac{2-\gamma}{N}}.
\end{eqnarray}
where 
\begin{eqnarray}
m^B_{\gamma,n}&=&-j^B+n-\gamma.
\end{eqnarray}
Interestingly, the suppression $O(1/N)$ in the pseudo-inner product between states with different parity has a statistical origin encapsulated in the Clebsch-Gordan coefficients.
We can then obtain, assuming the chemical potential to be sufficiently low
\begin{equation}
\label{eq:sol_full}
\Gamma^{G_N\rightarrow {B'_{N-1}}}=\Gamma_\text{el} \left[g_N \frac{\omega_0 (\omega_c \cos\theta_{B'} + g_N \sin\theta_{B'}) }{ (\omega_0+\omega_c)(\omega_0 \omega_c-g_N^2)}\right]^2,
\end{equation}
with $\tan{\theta_+}=({-\Delta+\sqrt{4g_N^2+\Delta^2}})/(2g_N)$, given by \Eq{tanp}, and 
$\theta_+=\theta_{-}-\pi/2$ with $\Delta=\omega_0-\omega_c$, and $g=g_N=\chi\sqrt{N}$.
This expression for the single-polariton rates is plotted as a function of the cavity-matter detuning, in Figure~\ref{figappfermionsingle} valid in the fermionic case (dot-dashed curves), where it is compared with the corresponding estimates obtained in the full bosonic model (dashed curves) and perturbative bosonic model (solid curves), showing an excellent quantitative agreement between the three models up to large detuning, $|\omega_c-\omega_0|=0.5 \omega_0$. The values for the plot is $g=g_N=0.1\omega_0$, with $I \equiv \Gamma_\text{el}$.
\subsection{Model comparisons and limitations}
\label{sec:appGl}
In order to ascertain the validity of the perturbative bosonic approximation, we derived the full bosonic Hamiltonian (non-perturbative in light-matter coupling) and we developed a fully fermionic theory, which allows to retain the results for the double-polariton states. 
To test the quantitative predictions gained with such simple and intuitive bosonic model, we have also developed a full second-quantization fermionic theory which clearly shows that the results hold even if this more refined approach is employed.
Both approaches are in good qualitative and quantitative agreement with the perturbative bosonic theory.

The fermionic transport analysis performed here is not limited to solid-state semiconductor devices, but it can be mapped to an open Dicke model with scattering. However, in that setting, which has so far been realized experimentally in lattices of trapped atoms, it could be challenging to engineer and modulate a process equivalent to the electron current we describe herein.

For simplicity, we have here assumed that the electron and photonic reservoirs were unstructured. It is an open question how ground-state electroluminescence could be affected by structured baths. Our calculations show that the effect is visible already in a dilute-current regime, yet at higher current flows, more complex effects could arise. The effect of electron-electron scattering, a second-order process in the dilute regime could also be addressed. The effect of a multi-mode model for the cavity might give rise to nonlinear quantum phenomena.

%\end{widetext}
%%%
%%%
\end{document}